\documentclass[sigplan,10pt,screen,nonacm]{acmart}
\renewcommand\footnotetextcopyrightpermission[1]{}

\usepackage{multirow}
\usepackage{tcolorbox}
\usepackage{comment}
\tcbuselibrary{breakable,skins}
\usepackage{placeins} 

\usepackage{natbib}
\usepackage[utf8]{inputenc} 
\usepackage[T1]{fontenc}    
\usepackage{hyperref}       
\usepackage{url}            
\usepackage{booktabs}       
\usepackage{amsfonts}       
\usepackage{nicefrac}       
\usepackage{microtype}      
\usepackage{xcolor}         
\usepackage{xspace}
\usepackage{amsmath}
\usepackage{caption}
\usepackage{ulem}

\usepackage{listings} 

\usepackage{svg}

\usetikzlibrary{arrows.meta,positioning,fit,calc,shapes.geometric}

\newcommand{\eat}[1]{} 

\newcommand{\after}[1]{[{\color{red}AFTER: #1}]}
\renewcommand{\after}[1]{} 

\newcommand{\ie}{{\em i.e.}, }

\newcommand{\heading}[1]{\vspace{.5pt}\noindent\textbf{#1}:\enspace}

\usepackage{minted}
\lstdefinestyle{mystyle}{
    backgroundcolor=\color{backcolour},   
    commentstyle=\color{codegreen},
    keywordstyle=\color{magenta},
    numberstyle=\tiny\color{codegray},
    stringstyle=\color{codepurple},
    basicstyle=\ttfamily\footnotesize,
    breakatwhitespace=false,         
    breaklines=true,                 
    captionpos=b,                    
    keepspaces=true,                 
    numbers=left,                    
    numbersep=5pt,                  
    showspaces=false,                
    showstringspaces=false,
    showtabs=false,                  
    tabsize=2
}

\newcommand{\sys}{\mbox{\textsc{VineLM}}\xspace}

\newcommand{\LLMSet}{{\mathcal L}}
\newcommand{\QuerySet}{{\mathcal Q}}

\usepackage{enumitem}
\setlist[itemize]{leftmargin=*}

\begin{document}

\date{}

\title{\sys: Trie-Based Fine-Grained Control for\\Agentic Workflows}

\author{Nikos Pagonas}
\affiliation{%
  \institution{Columbia University}
  \country{}
}
\email{n.pagonas@columbia.edu}

\author{Matthew Lou}
\affiliation{%
  \institution{Columbia University}
  \country{}
}
\email{ml4855@columbia.edu}

\author{Tianyi Peng}
\affiliation{%
  \institution{Columbia University}
  \country{}
}
\email{tianyi.peng@columbia.edu}

\author{Dan Rubenstein}
\affiliation{%
  \institution{Columbia University}
  \country{}
}
\email{danr@cs.columbia.edu}

\author{Kostis Kaffes}
\affiliation{%
  \institution{Columbia University}
  \country{}
}
\email{kkaffes@cs.columbia.edu}

\begin{abstract}
Agentic workflows interleave configurable LLM stages with tool stages and often include retries or refinement loops.
Existing workflow managers profile full workflow configurations offline and assign each request a static workflow-level plan that binds each configurable LLM stage to a single model, reuses that model across repeated loop iterations, and does not revisit those choices at runtime.
We present \sys, a workflow manager that enables fine-grained control by choosing the model for each stage invocation as execution unfolds under request-level objectives such as maximizing accuracy under cost or latency budgets.
\sys represents feasible executions as an annotated trie of model-choice prefixes and uses checkpointing and cascade profiling to estimate path accuracy, cost, and latency without exhaustively profiling every request on every path.
At runtime, \sys re-roots the trie after each stage invocation and replans over the remaining subtrie using the realized execution prefix and remaining latency budget.
On NL2SQL and math reasoning workflows, \sys improves the cost-latency-accuracy frontier over coarse workflow-level baselines, achieving up to 18\% higher accuracy at the same per-request budget with its sparse profiling reducing offline profiling cost by 98--99.8\% when compared to exhaustive profiling.
\end{abstract}

\settopmatter{printfolios=true}
\pagestyle{plain}
\maketitle

\section{Introduction}
\label{sec:intro}

\heading{Agentic workflows}
Modern AI applications increasingly take the form of agentic workflows rather than a single monolithic LLM call~\cite{hugginggpt,metagpt,agent_survey,react}.
A workflow is a stateful program composed of stages: some stages are \emph{configurable LLMs}, while others are {\em tools} such as retrieval, SQL execution, or external API calls~\cite{toolLLM,toolformer,helium,gorilla,rewoo}.
A request is handled by a sequence of stage invocations, and intermediate results determine what stage, if any, happens next, allowing the workflow to retry, refine, and maintain state across steps.
A workflow is conceptually the analog of a program where stages are the instructions.  Like programs, workflows can conceptually implement branching and looping~\cite{aflow,agentxray,treeofthoughts,graphofthoughts,planandsolve}.
Frameworks such as LangGraph~\cite{langgraph}, LlamaIndex Workflows~\cite{llamaindex}, AG2~\cite{ag2}, among others~\cite{crewai,openai_agents_sdk,strands_agents,claude_agent_sdk,lmql,dspy} already expose this programming model.

\begin{figure}[t]
    \centering
    \includegraphics[width=1\linewidth]{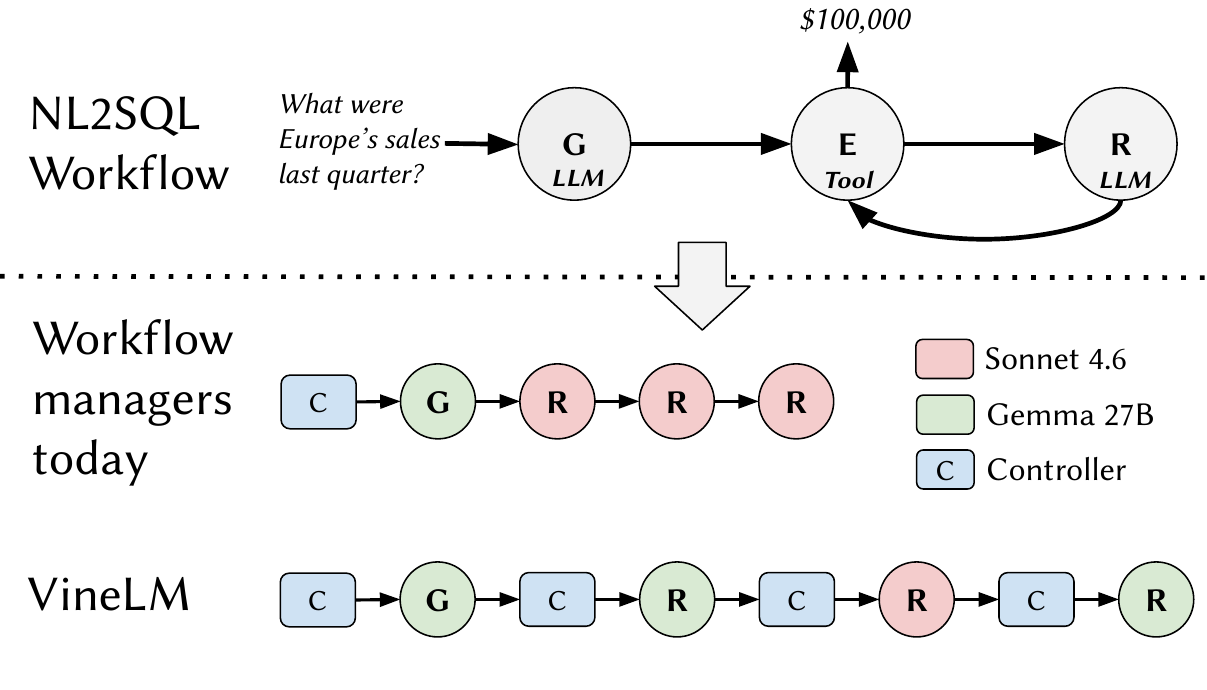}
    \caption{A simplified example NL2SQL workflow containing a SQL Generation LLM stage (G), a SQL Execution tool stage (E), and a potentially repeated Repair LLM stage (R). Existing workflow managers such as Murakkab~\cite{chaudhry2025murakkab} commit to one workflow-level model assignment when the request arrives, with repeated loop iterations always reusing the same model. \sys instead reconsiders the choice after each stage and can mix LLMs across different iterations of the same loop. This finer-grain control yields a much larger decision space, but enables better end-to-end optimization.}
    \label{fig:nl2sql}
\end{figure}

A representative example is NL2SQL shown in Figure~\ref{fig:nl2sql}.
Given a question such as ``What were Europe’s sales last quarter?'', the workflow generates a candidate SQL query, executes it, and uses execution feedback (e.g., runtime error) or available offline information (e.g., I need two iterations to reach my target average accuracy) to decide whether to stop or continue in a bounded repair loop.
Generation and Repair are configurable LLM stages, while SQL Execution is a tool stage.
This feedback-and-repair pattern is now common well beyond NL2SQL, from self-refinement systems~\cite{madaan2023selfrefine} to tool-using agents that improve through execution feedback and retry~\cite{shinn2023reflexion}.

\heading{Workflow management today}
Managing such workflows requires choosing a {\em model} for each configurable LLM stage, where models differ in cost, latency, and accuracy.
State-of-the-art managers such as Murakkab~\cite{chaudhry2025murakkab} profile a space of full \emph{workflow configurations} offline, where each configuration binds a specific model to each configurable stage in the workflow template.
Crucially, this binding is to the stage template, not to each dynamic stage invocation.
If a workflow contains a loop, the manager does not unroll that loop into separate first-, second-, and third-iteration decisions; instead, it chooses one model for the stage and must reuse that same model on every loop iteration within the request.
At serving time, each request is therefore assigned one static configuration before the workflow begins.
In the NL2SQL example of Figure~\ref{fig:nl2sql}, the system can choose one model for Generation and a different model for Repair, but it cannot use intermediate execution feedback to switch to a different model on later repair rounds.
We refer to this decision granularity as \emph{coarse-grained control}: the manager chooses one workflow-level model assignment up front and keeps that assignment fixed for the lifetime of the request.

As we show in \S\ref{sec:motivation}, coarse-grained control can severely limit performance.
In refinement loops, the best model for an early repair attempt is often not the best model for a later one, and mixed-model trajectories can dominate plans that bind a single model to the repair stage for all iterations.
Runtime variability makes static plans even weaker: a downstream choice that looked best at the start of the request can become the wrong one after an unexpectedly slow or expensive earlier stage.
Current systems stop at offline workflow-level plans to avoid combinatorial explosion of the space to be considered.
Once each refinement step is allowed to choose its model independently, the number of feasible paths grows rapidly.
For example, in our NL2SQL workflow, with 8 candidate models at Generation and at each of two Repair rounds, the workflow already induces 584 feasible model-choice paths.
Exhaustively profiling that space is expensive even before considering online adaptation.


\heading{Our approach}
In contrast, this paper targets \emph{fine-grained control}: choosing the model separately for each stage invocation as the workflow unfolds.
Under fine-grained control, the first Repair invocation and the second Repair invocation are distinct decision points, even though they correspond to the same logical Repair stage in the workflow template.

Our key insight is to represent this space as an \emph{execution trie}.
Each node corresponds to a prefix of model choices for the stage invocations encountered so far in a workflow run, including repeated invocations of the same logical stage inside a refinement loop.
Thus, every root-to-node path represents a feasible workflow prefix: if the workflow can terminate at that node, it is a complete feasible run; otherwise, it is a partial run that can be extended to one of the node's descendants.
The trie exposes shared prefixes across many candidates, turning a large set of full plans into a structured search space.
We show that this structure helps in two ways.
First, it enables \emph{efficient offline annotation}: using checkpointing, cascade-style profiling, and subtree reuse, we can estimate the accuracy, cost, and latency of many prefixes and terminating runs from a small number of sampled executions.
Second, it enables \emph{efficient online adaptation}: after each stage invocation completes, we can re-root the trie at the realized prefix, revise latency to reflect actual time spent thusfar, and search only the remaining subtrie, allowing \sys\ to adapt later model choices to the observed progress of the request and, when available, the current system state.

We build \sys, an agentic workflow manager.
\sys constructs an annotated trie offline from representative traces and uses it online to perform per-invocation model selection under request-level objectives such as maximizing accuracy under a budget or minimizing cost subject to an accuracy floor.
Rather than commit to a single workflow-level plan at request start, \sys interleaves execution and control throughout the workflow run.
After each stage invocation completes, as depicted in Figure~\ref{fig:nl2sql}, the \sys controller uses its offline estimates together with the latency so far to choose the model for the next stage invocation.

We evaluate \sys on NL2SQL and math reasoning workflows.
Our results show three main takeaways.
First, per-invocation control materially improves the cost--latency--accuracy frontier over coarse workflow-level baselines, achieving up to 18\% higher accuracy with the same available models and the same workflow under the same budget.
Second, sparse cascade profiling annotates the trie accurately enough for control while requiring only 0.2\%--2\% of the cost of exhaustive profiling.
Third, rerooting after each stage reduces latency-SLO violations relative to static workflow-level plans by up to 85\%.

In summary, this paper makes four contributions:
\begin{itemize}
    \item We show that coarse-grained control leaves performance on the table in agentic workflows.
    \item We formulate fine-grained workflow control as search over an annotated execution trie of workflow prefixes.
    \item We design and implement \sys, which combines the trie formulation with checkpointing, efficient offline estimation, and dynamic rerooting to enable fine-grained per-invocation control.
    \item We demonstrate on NL2SQL and math reasoning workflows that \sys improves the cost-latency-accuracy frontier and reduces latency-SLO violations relative to baselines with very low profiling cost.
\end{itemize}

\section{Motivation}
\label{sec:motivation}

Our work here is motivated by the observation that existing workflow system controls are {\em coarse-grained,} in the sense that each LLM stage of a workflow binds to a fixed model.
We show in this paper that there are clear advantages to a more {\em fine-grained}, flexible controller whose assignment of a model to a stage can vary across iterations of that stage.  This is the case even when multi-iteration assignment is offline (a priori planned per stage iteration), but especially when assignment is online (i.e., selected on-the-fly during the execution of the workflow).

A representative design of a coarse-grained control system, exemplified by Murakkab~\cite{chaudhry2025murakkab}, profiles a set of candidate workflow configurations offline---e.g., one model assignment per configurable stage together with a fixed retry horizon---and then, at serving time, assigns each request one of those pre-profiled configurations.
This is an important step forward, but it is not the same as the fine-grained control we target.
The chosen configuration fixes the model assignment of each stage template.
In particular, if a workflow contains a refinement loop, the manager does not treat the each visit to that stage as a separate decision point; instead, it chooses one model up front and reuses that same model on every loop iteration within the request.
Any adaptation therefore happens at the level of switching among whole-workflow configurations, not at the level of choosing the next model after each stage invocation.
The system reasons about the workflow as one configurable unit, rather than as a sequence of stage invocations whose best next model can change as execution unfolds.

This distinction is the starting point for our motivation.
Below, we show that this coarse-grained control leaves performance on the table in two ways: it cannot mix models for a given stage across loop iterations, and it cannot revisit downstream choices after earlier stages consume unexpected (more or less) cost or latency budget than anticipated.

\subsection{Fixed loop choices}
\label{motivation:loop}
Coarse-grained control is especially restrictive in workflows with refinement loops.
In an NL2SQL workflow, a manager such as Murakkab profiles configurations of the form ``Generation = $L_i$, Repair = $L_j$, allow up to $h$ repairs,'' then assigns one such configuration to the request.
This lets the system choose one model for Generation and one model for Repair, but every visit to the Repair stage must reuse that same Repair model.
The first and second Repair invocations are therefore not treated as separate decision points, even though they arise after different failures and under different remaining budgets.

\begin{figure}[t]
    \centering
    \includegraphics[width=0.9\linewidth]{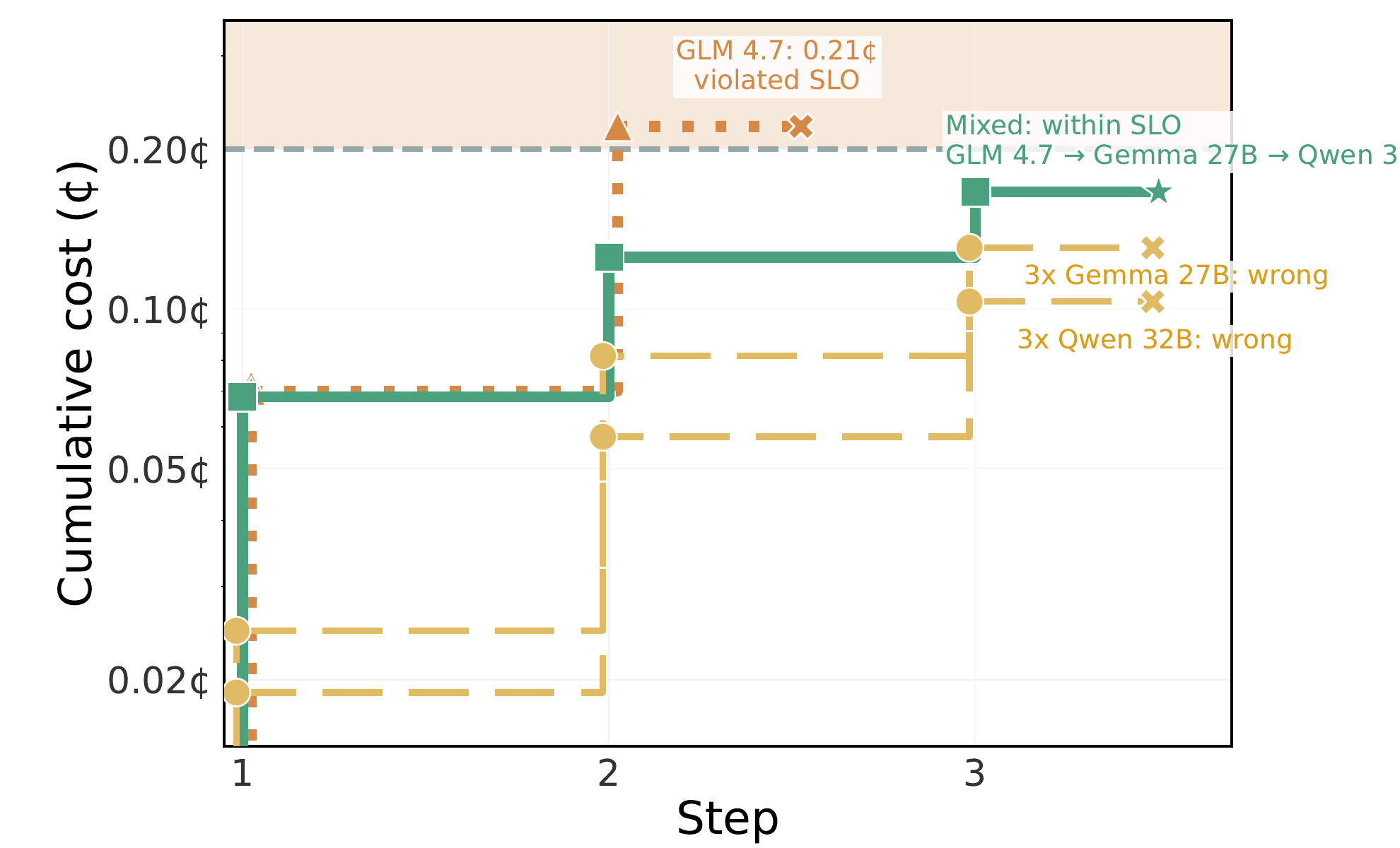}
    \caption{For a specific NL2SQL request, the objective is to return the correct answer under a fixed cost SLO. Static single-model plans fail in opposite ways: always using GLM 4.7 reaches the correct answer but exceeds the budget, while always using Gemma 27B or always using Qwen 32B stays within budget but returns the wrong answer. A mixed path---GLM 4.7 for generation, then Gemma 27B and Qwen 32B on successive repair rounds---stays within the SLO and returns the correct answer.
    This illustrates why refinement loops should be optimized as a sequence of model choices, not by binding one model to the entire loop.}
    \label{fig:motivation:mixed-models}
\end{figure}

Figure~\ref{fig:motivation:mixed-models} shows why this matters on a concrete NL2SQL request with a fixed cost SLO.
Static single-model strategies fail in opposite ways: always using \texttt{GLM 4.7} reaches the correct answer but exceeds the budget, while always using \texttt{Gemma 27B} or always using \texttt{Qwen 32B} stays within budget but returns the wrong answer.
The successful path is mixed:
\texttt{GLM 4.7} for the initial generation, then \texttt{Gemma 27B} and \texttt{Qwen 32B} on successive repair rounds.
The figure uses the simplest static alternatives to make the effect visually clear, but the underlying point is broader: any coarse workflow-level plan that binds one model to the Repair stage cannot realize this successful path, because it requires different model choices on different loop iterations.

More generally, the right object to optimize is the sequence of model choices across loop iterations, not a single model bound to the loop as a whole.
An early repair may be best served by a cheaper model that quickly resolves easy cases, while a later repair may justify escalating to a stronger model because the marginal value of avoiding another failure is now higher.
Coarse-grained control cannot express such trajectories because it never treats individual loop iterations as separate decision points.

\subsection{No per-invocation adaptation}
Coarse-grained control fixes more than the model reused within a refinement loop: it fixes the remaining workflow plan before execution reveals how the current request is actually progressing.
Offline profiling can identify the workflow-level configuration with the best expected cost--latency--accuracy tradeoff for a target objective, but that choice is based on averages across prior runs.
For any particular request, the realized latency of an invocation depends not only on the chosen model, but also on the request itself, the amount of generated text, and transient backend conditions while the request is in flight~\cite{amdahltail,tailatscale}.
As a result, a configuration that looks optimal at request start can become suboptimal---or even infeasible---after one unexpectedly slow step.
Existing workflow managers can react to load by periodically refreshing workflow-level configurations or by selecting among pre-profiled ones, but they do not reconsider the next model after each stage invocation~\cite{chaudhry2025murakkab}.

\begin{figure}[t]
    \centering
    \includegraphics[width=0.9\linewidth]{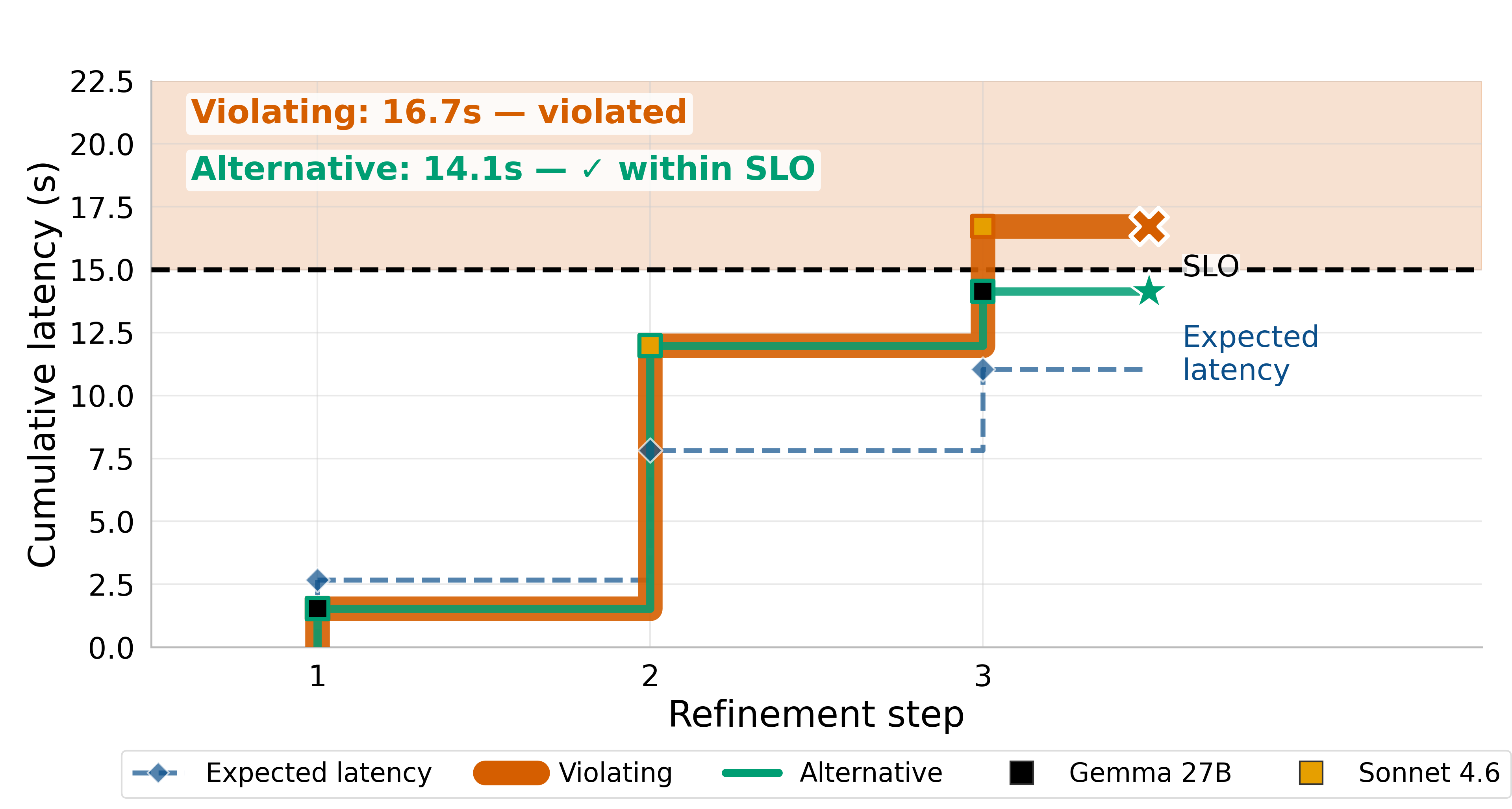}
    \caption{For this MathQA request with three stages, the objective is to maximize accuracy while meeting a 15\,s latency SLO. Offline averages favor the plan \texttt{Gemma 27B} $\rightarrow$ \texttt{Sonnet 4.6} $\rightarrow$ \texttt{Sonnet 4.6}, but the realized second step runs long, so following that plan finishes in 16.7\,s and violates the SLO (orange). Replanning after step 2 and switching the final step to \texttt{Gemma 27B} yields \texttt{Gemma 27B} $\rightarrow$ \texttt{Sonnet 4.6} $\rightarrow$ \texttt{Gemma 27B}, finishing in 14.1\,s and staying within the SLO (green).}
    \label{fig:motivation:online-is-needed}
\end{figure}

Figure~\ref{fig:motivation:online-is-needed} shows a concrete example on a self-reflective MathQA workflow for a request with a 15\,s latency SLO and the objective of maximizing accuracy.
Using offline average profiles, the best workflow-level plan is
\texttt{Gemma27B} $\rightarrow$ \texttt{Sonnet 4.6} $\rightarrow$ \texttt{Sonnet 4.6}.
At runtime, however, the second step---the first \texttt{Sonnet 4.6} invocation---takes longer than expected.
If the system continues with the original plan, the request exceeds the 15\,s budget.
A controller with per-invocation adaptation can instead replan after observing that delay and switch the final step to \texttt{Gemma27B}, yielding
\texttt{Gemma27B} $\rightarrow$ \texttt{Sonnet 4.6} $\rightarrow$ \texttt{Gemma27B},
which remains within the latency SLO.

The key point is that the best remaining suffix depends on the realized execution prefix, not just on offline averages.
Thus, what is missing is fine-grained adaptation within a workflow run.
If an early stage has already consumed too much latency budget, the system should be able to switch the next Repair invocation to a faster model or to a shorter remaining refinement strategy.
If earlier stages finish quickly, it should be able to spend the saved slack on a stronger downstream model where additional accuracy is most valuable.

\section{Trie-Based Workflow Formulation}
\label{sec:formulation}

\subsection{Setting, scope, and request objectives}
\label{sec:formulation:setting}

We consider a fixed agentic workflow template $W$ that may contain branches and refinement loops. Some stages in $W$ are tool stages, such as a SQL engine or retrieval component, while others are \emph{configurable} LLM stages at which the system may choose among multiple models. Let
\[
\LLMSet = \{L_1, \ldots, L_m\}
\]
be the set of available LLMs. Each model $L_i$ has its own quality, latency, and monetary cost characteristics, and each configurable stage $s$ in the workflow admits a subset $\LLMSet(s) \subseteq \LLMSet$ of valid model choices. We are also given an offline dataset of representative requests
\[
\QuerySet = \{q_1, \ldots, q_n\},
\]
together with ground truth (i.e., the accurate output) for each request. 

Our goal is to support \emph{request-specific} objectives over accuracy, latency, and cost. Each incoming request is therefore modeled as a pair $(q,o)$, where $q$ is the request itself and $o$ is an objective specification. We write
\[
o = (f, \mathcal{C}),
\]
where $f$ is the quantity to optimize and $\mathcal{C}$ is a set of constraints. Typical objectives include:
\[
\begin{aligned}
&\text{minimize cost}      &&\text{subject to accuracy } \ge a,\\
&\text{maximize accuracy}  &&\text{subject to cost } \le c,\\
&\text{maximize accuracy}   &&\text{subject to latency } \le l.
\end{aligned}
\]
This formulation and our approach is also easily extended to optimizations of one criteria subject to constraints on the other 2, e.g., maximize accuracy subject to both cost $\le c$ and latency $\le l$.

A key feature of our setting is that these targets are \emph{absolute} and may vary from request to request; we do not assume a fixed global SLO or a small set of coarse quality tiers.

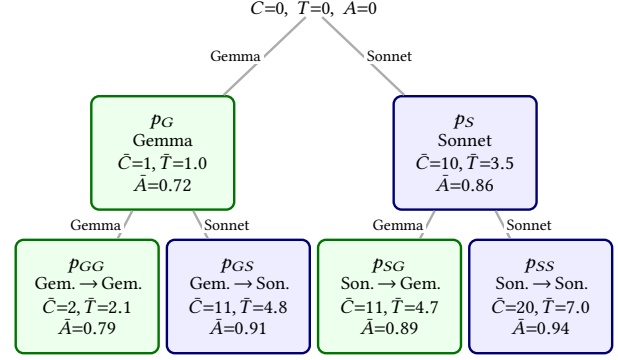
\begin{figure}[t]
\centering
\newcommand{\bC}{\smash{\bar C}}
\newcommand{\bT}{\smash{\bar T}}
\newcommand{\bA}{\smash{\bar A}}
\begin{tikzpicture}[
  level distance=19mm,
  level 1/.style={sibling distance=40mm},
  level 2/.style={sibling distance=20mm},
  edge from parent/.style={draw=gray!65, line width=0.9pt},
  every node/.style={font=\small},
  state/.style={
    draw,
    rounded corners=3pt,
    line width=0.9pt,
    align=center,
    inner sep=2pt,
    text width=17.5mm,
    minimum height=15mm,
    font=\scriptsize
  },
  gemma/.style={state, fill=green!7, draw=green!45!black},
  sonnet/.style={state, fill=blue!7, draw=blue!45!black},
  elabel/.style={midway, fill=white, inner sep=1pt, font=\tiny}
]
\node[align=center, inner sep=0pt, font=\scriptsize]
  {$\bC{=}0,\;\bT{=}0,\;\bA{=}0$}
child {
  node[gemma] {$p_G$\\[0pt] Gemma\\[1pt]
    $\bC{=}1,\bT{=}1.0$\\$\bA{=}0.72$}
  child {
    node[gemma] {$p_{GG}$\\[0pt]
      Gem.\,$\to$\,Gem.\\[1pt]
      $\bC{=}2,\bT{=}2.1$\\$\bA{=}0.79$}
    edge from parent node[elabel,left] {Gemma}
  }
  child {
    node[sonnet] {$p_{GS}$\\[0pt]
      Gem.\,$\to$\,Son.\\[1pt]
      $\bC{=}11,\bT{=}4.8$\\$\bA{=}0.91$}
    edge from parent node[elabel,right] {Sonnet}
  }
  edge from parent node[elabel,left] {Gemma}
}
child {
  node[sonnet] {$p_S$\\[0pt] Sonnet\\[1pt]
    $\bC{=}10,\bT{=}3.5$\\$\bA{=}0.86$}
  child {
    node[gemma] {$p_{SG}$\\[0pt]
      Son.\,$\to$\,Gem.\\[1pt]
      $\bC{=}11,\bT{=}4.7$\\$\bA{=}0.89$}
    edge from parent node[elabel,left] {Gemma}
  }
  child {
    node[sonnet] {$p_{SS}$\\[0pt]
      Son.\,$\to$\,Son.\\[1pt]
      $\bC{=}20,\bT{=}7.0$\\$\bA{=}0.94$}
    edge from parent node[elabel,right] {Sonnet}
  }
  edge from parent node[elabel,right] {Sonnet}
};
\end{tikzpicture}
\caption{Illustrative execution trie with two configurable stages and two admissible models per stage. Each node $p$ is annotated with expected cumulative cost $\bar C(p)$, expected cumulative latency $\bar T(p)$, and expected accuracy $\bar A(p)$ if execution terminated at that prefix. All three metrics are monotone along each root-to-leaf path.}
\label{fig:formulation:toy_trie}
\end{figure}

\subsection{Workflow templates and execution tries}
\label{sec:formulation:trie}

A \emph{workflow path} is a sequence of model choices at the configurable stages encountered during one realization of the workflow:
\[
p = (\ell_1, \ell_2, \ldots, \ell_k), \quad \ell_i \in \LLMSet(s(i))
\]
where $s(i)$ is the $i$th stage of path $p$.
Different paths may have different lengths $k$ depending on how many configurable stages are traversed. The same model may appear at multiple positions due to loops.  We assume each loop is bounded by a maximum retry horizon so that the set of feasible execution paths is finite.

The set of all feasible workflow paths induces an \emph{execution trie}. The root corresponds to the empty prefix. Each node corresponds to a prefix of model assignments,
\[
u = (\ell_1, \ldots, \ell_r),
\]
and there is an edge from $u$ to
\[
u' = (\ell_1, \ldots, \ell_r, \ell_{r+1})
\]
whenever appending the next model choice is legal under the workflow's control flow and loop bounds. A root-to-leaf path represents one complete workflow instance, while an internal node represents a partial execution prefix shared by many complete paths. Unlike a full $|\LLMSet|$-ary tree, this trie respects workflow semantics: not every model is valid at every position, and some branches terminate earlier than others.

This trie view is useful because it makes the combinatorial structure of the decision space explicit while also exposing shared prefixes across many candidate executions. Two workflow instances that make the same early choices but diverge later share the same prefix nodes, and repeated loop iterations appear naturally as deeper prefixes of the same stage family. In the ideal case, each node in the trie would be annotated with the cumulative latency, expected cost, and expected accuracy associated with executing the corresponding prefix. The next subsection formalizes these path metrics and shows how monotonicity induces a structured feasible region for request-level optimization.

\subsection{Path metrics}
\label{sec:formulation:metrics}
For the remainder of this section, we abuse notation slightly and use $p$ to denote both a node in the execution trie and the workflow plan induced by the model assignments on the root-to-$p$ path.
Descendants of $p$ correspond to strictly longer workflow plans that allow additional refinement rounds.

Let $p=(\ell_1,\ldots,\ell_k)$, and let $R_i(q,p)$ indicate that the $i$th stage on that prefix is actually reached when serving request $q$ under the workflow's normal stop conditions.
For each request $q \in \QuerySet$ and workflow path $p$, we define three end-to-end metrics:
\[
A(q,p), \qquad C(q,p), \qquad T(q,p),
\]
corresponding to task accuracy, monetary cost, and latency.

In the simplest case, $A(q,p)\in\{0,1\}$ is an offline success indicator computed against the request's ground truth.
Monetary cost is treated in expectation:
\[
C(q,p) = \sum_{i=1}^{k} R_i(q,p)\, c_i(q,\ell_i),
\]
where $c_i(q,\ell_i)$ is the realized dollar cost of stage $i$ if it is executed.
Thus, if the workflow terminates early at an ancestor of $p$, all later stages on the prefix contribute zero cost.
The node annotation
\[
\bar C(p)=\mathbb{E}_{q\sim\mathcal D}[C(q,p)]
\]
therefore already accounts for early termination and should be read as the \emph{expected spend} of choosing prefix $p$.

Latency is handled more conservatively.
Because cost budgets are naturally expectation-based but latency SLOs are per-request constraints that should not be violated by continuing too deep into the workflow, we do not discount later-stage latency by the probability of early stopping when annotating a node.
Instead, if $\tau_i(q,\ell_i)$ denotes the latency of stage $i$ when it is executed, we define the node's latency annotation as
\[
\bar T(p) = \sum_{i=1}^{k} \mathbb{E}_{q\sim\mathcal D}\!\left[\tau_i(q,\ell_i)\mid R_i(q,p)=1\right],
\]
that is, the sum of expected per-stage latencies along the prefix.
Intuitively, a request that continues to a later stage must still have enough remaining wall-clock budget to execute that stage; discounting its latency by the probability of early termination would understate the time required on the requests that actually continue.

Finally,
\[
\bar A(p)=\mathbb{E}_{q\sim\mathcal D}[A(q,p)].
\]
Figure~\ref{fig:formulation:toy_trie} illustrates a small execution trie annotated with $(\bar C,\bar T,\bar A)$ at each node.

\subsection{Oracle path selection}
\label{sec:formulation:oracle}

Our approach relies on the observation that when the offline set of query samples $\QuerySet$ is sufficiently large, our mean estimates of $A(p), C(p),$ and $T(p)$ accurately describe the means of future query samples.  Specifically, when the $q \in \QuerySet$ form a distribution that is stationary the law of large numbers ensures that  a sufficiently large set of subsequent alternate online queries $\QuerySet'$ would, if similarly measured offline, yield identical mean esimates.
With this assumption, if the execution trie were fully annotated with exact path metrics $\bar{A}, \bar{C}, \bar{T}$, request handling would reduce to a constrained search over workflow paths. Let $\mathcal{P}$ denote the set of terminating nodes in the trie. A request objective $o = (f, \mathcal{C})$ specifies an optimization target $f$ and constraints $\mathcal{C}$. Three representative objectives are:

\heading{Minimize cost subject to accuracy floor}
\[
p^\star = \arg\min_{p \in \mathcal{P}}\ \bar C(p)
\quad\text{s.t.}\quad \bar A(p) \ge a.
\]
In Figure~\ref{fig:formulation:toy_trie}, for $a = 0.90$, only $p_{GS}$ ($\bar A=0.91$) and $p_{SS}$ ($\bar A=0.94$) satisfy the accuracy floor; the optimizer selects $p_{GS}$ because $\bar C(p_{GS})=11 < 20=\bar C(p_{SS})$.

\heading{Maximize accuracy subject to latency cap}
\[
p^\star = \arg\max_{p \in \mathcal{P}}\ \bar A(p)
\quad\text{s.t.}\quad \bar T(p) \le t.
\]
For $t = 5.0$ in the figure, all four terminating paths satisfy the cap, and $p_{SS}$ ($\bar A=0.94$) is selected.

\heading{Maximize accuracy subject to cost budget}
\[
p^\star = \arg\max_{p \in \mathcal{P}}\ \bar A(p)
\quad\text{s.t.}\quad \bar C(p) \le c.
\]
For $c = 11$, the feasible paths are $p_{GG}$, $p_{GS}$, and $p_{SG}$; the optimizer selects $p_{GS}$ ($\bar A=0.91$) as the most accurate within budget. These examples show that the optimal path is \emph{not} fixed: it depends on the request objective and the induced feasible region in the trie.

\heading{Remark (Monotonicity enables pruning)} One observation to speed up the above search is that all three metrics are monotone along any root-to-leaf path ($\bar A(u)\le\bar A(v)$, $\bar C(u)\le\bar C(v)$, $\bar T(u)\le\bar T(v)$ whenever $u$ prefixes $v$): cost and latency grow because descendants run strictly more stages; we also assume accuracy cannot decrease because additional stages only add refinement opportunities. Consequently, each feasibility constraint defines a contiguous boundary on any root-to-leaf chain, allowing subtrees that cannot contain feasible or optimal descendants to be pruned.

\heading{Remark (Online Adaptation)} The path selected above is an offline recommendation based on expected metrics. Online, if intermediate stage outcomes deviate from expectation (e.g. a stage invocation finishes sooner than expected), the system can re-evaluate the remaining subtrie and switch to a different path. We explore this setting in \S\ref{sec:design:dynamic}.

\subsection{Challenges in estimating the annotated trie}
\label{sec:formulation:challenges}
The oracle formulation is conceptually simple once the trie is fully annotated.
In practice, obtaining those annotations is the central challenge.

\heading{Scale} If we index requests by rows and workflow paths by columns, the accuracy outcomes form a \emph{request--path table}
\[
A \in \{0,1\}^{|\QuerySet| \times |\mathcal{P}|},
\]
with analogous tables $C$ and $T$ for cost and latency. In our NL2SQL experiments, $|\QuerySet| = 1{,}529$ and $|\mathcal{P}| = 584$ ($8 + 64 + 512$ paths at depths 1--3), giving a $1{,}529 \times 584$ table with over 890{,}000 entries. Fully observing $A$ would require running every request on every path — infeasible even for moderate workflow sizes. The central problem is therefore to estimate $A$ under a limited profiling budget.

This is naturally viewed as a matrix completion problem~\cite{candes2009matrixcompletion,koren2009matrixfactorization,softimpute}: we observe a sparse subset of entries and wish to estimate what is missing. Our setting, however, differs from standard matrix completion in two important ways.

\heading{Column mean estimation, not entry recovery} Standard matrix completion aims to recover every entry of a low-rank matrix. Here we only need the column means $\bar A(p) = \frac{1}{|\QuerySet|}\sum_q A(q,p)$ for path selection. This relaxed goal allows estimators that exploit the known cascade structure directly without attempting to recover the full matrix.

\begin{figure}[t]
\centering
\includegraphics[width=0.9\linewidth]{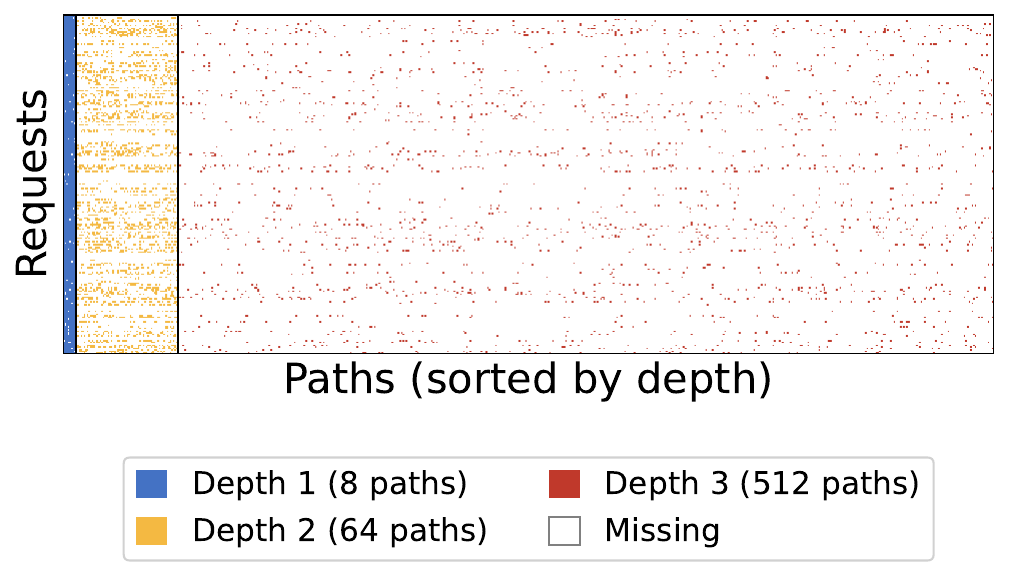}
\caption{The request--path accuracy table $A$ under 5\% cascade sampling. Columns are sorted left to right by depth: 8 depth-1 paths, 64 depth-2 paths, and 512 depth-3 paths (separated by vertical lines). Missingness is severe and depth-dependent: depth-1 columns are 97\% observed, depth-2 only 17\%, and depth-3 just 1.8\%, because later stages are only reached when earlier stages are executed. Each color indicates a depth-level observed entry; white indicates missing.}
\label{fig:mnar_matrix}
\end{figure}

\heading{Missing Not At Random (MNAR) observation pattern} Our approach to sampling paths is for each query, to repeatedly  pick a random node on the trie.  We make 2 observations: a depth $d$ node can be executed when either it is explicitly chosen or when one of its descendants is chosen.  Also, the number of depth $d$ nodes increases monotonically in $d$, reducing the likelihood of sampling a specific depth $d$ node than depth $d'<d$ counterparts.  Figure~\ref{fig:mnar_matrix} shows the result: depth-1 columns are 97\% observed, depth-2 only 17\%, and depth-3 just 1.8\%. This is not only a sparsity issue: the accuracy monotonicity property (\S\ref{sec:formulation:metrics}) also imposes additional constraints on $A$. Typical matrix completion does not exploit these structures. 

\heading{Estimation for optimization} The goal of offline profiling is not merely to minimize estimation error uniformly. Rather, the estimates serve downstream path selection: small errors near a feasibility boundary can flip the selected path, while larger errors far from the boundary are inconsequential. 

Taken together, the execution trie, MNAR estimation under a limited profiling budget, and downstream path optimization form a rich problem space. The next section describes how \sys addresses these challenges.

\section{\sys Design}
\label{sec:design}

\subsection{Design Overview}
\label{sec:design:overview}

\begin{figure}[t]
\centering
\includegraphics[width=0.47\textwidth]{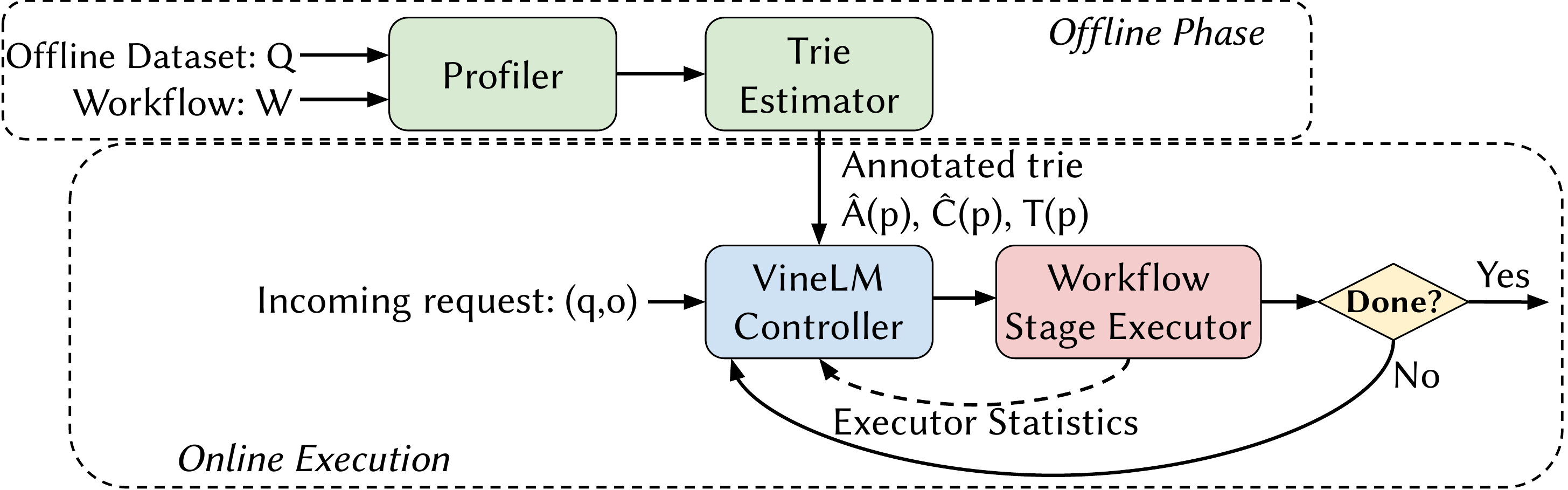}
\caption{\sys overview.}
\label{fig:system_overview}
\end{figure}

Figure~\ref{fig:system_overview} shows the architecture of \sys which has two main components: an \emph{offline trie construction pipeline} that builds an annotated execution trie, and a \emph{runtime controller} that uses that trie to make invocation-level decisions for each request. Offline, the developer provides a workflow template $W$ together with a representative dataset $\QuerySet$ for which ground truth is available. The profiler then evaluates a selected set of request--path pairs rather than exhaustively running every request on every workflow instance. These sampled outcomes are passed to the trie estimator, which constructs an annotated trie whose nodes correspond to execution prefixes and whose annotations estimate the expected metrics
${\bar A}(p),{\bar C}(p),{\bar T}(p)$
for each prefix or terminating path $p$. Intuitively, this trie is the compact summary that lets \sys\ reason about a large space of workflow executions without materializing every possible run in full.

At runtime, each incoming request arrives together with an objective $o=(f,\mathcal{C})$, such as minimizing cost subject to an accuracy floor and latency cap. The \sys\ controller begins at the root of the trie and repeatedly chooses the next stage/model decision based on the current execution prefix, the remaining budget implied by $o$, and the estimated metrics stored in the trie. The chosen action is then executed by the workflow stage executor, which may invoke either an LLM stage or a tool stage. If the workflow terminates, the result is returned to the user; otherwise, execution advances to a new trie prefix and the controller makes another decision. In this way, \sys\ does not commit to a single workflow-level plan at admission time. Instead, it interleaves execution and control, revising its choices as the request progresses through the workflow.

\subsection{Offline Trie Annotation}
\label{sec:design:build}

Building the annotated trie requires estimating, for each workflow path $p$, the expected path-level metrics
$\hat{\bar A}(p)$, $\hat{\bar C}(p)$, and $\hat{\bar T}(p)$.
In principle, all three must be learned from limited profiling data.
In practice, the hard part is accuracy.
Path-level cost and latency are largely determined by the chosen model, stage, and infrastructure, so their averages are comparatively stable and easier to estimate; moreover, \sys can observe realized cost and latency online and update the remaining budget after each stage.
We therefore focus on offline accuracy estimation in this subsection.

\heading{Checkpointing}
Many workflow paths share a long prefix.
If
\[
p_1=(\ell_1,\ldots,\ell_k), \qquad
p_2=(\ell_1,\ldots,\ell_b,\ell'_{b+1},\ldots,\ell'_{k'}),
\]
then the state after the common prefix $(\ell_1,\ldots,\ell_b)$ can be checkpointed once and reused to evaluate both suffixes.
This reduces profiling cost and also removes spurious variance from rerunning the shared prefix independently for each descendant.
Although reuse lowers the number of independent samples of the prefix itself, this can be offset by replaying a query along multiple balanced paths if needed.  Storage space to hold checkpointed an be constrained by ordering execution of the randomly selected queries that operate over similar parts of the trie (i.e., use the same checkpoints) successively.

\heading{Cascade sampling}
To populate the annotated trie under a limited profiling budget, \sys profiles the workflow in cascade order.
For each sampled query $q$, it invokes a randomly selected depth-1 model $\ell_1$; if $\ell_1$ fails, it continues to a depth-2 extension $(\ell_1,\ell_2)$; and so on until the query succeeds or the path is exhausted.
Cascade execution also yields free additional labels through \emph{subtree fill-in}.
Under our path semantics, if a query succeeds at some prefix, then every extension of that prefix is also counted as successful, since success anywhere on the path makes the whole path successful.
Thus, when a node succeeds, the entire subtree rooted at that node can be marked successful at no extra cost.
A budget of $N$ cascade runs therefore yields substantially more than $N$ observed request--path entries, with shallow columns effectively close to fully observed.

\heading{Why standard matrix completion remains biased}
Despite subtree fill-in, the resulting observation pattern is still MNAR.
Consider a depth-2 path $(\ell_1,\ell_2)$.
Cascade execution reaches this column only on queries for which $\ell_1$ failed, so the raw sample mean estimates
$\Pr[\ell_2 \text{ succeeds} \mid \ell_1 \text{ fails}]$,
not the target
\[
\bar A(\ell_1,\ell_2)=\Pr[\ell_1 \text{ or } \ell_2 \text{ succeeds}].
\]
More generally, deeper columns are observed on increasingly hard subpopulations.
A method that fits directly to these observed means will therefore systematically underestimate the true path-level accuracy of deeper paths.

\heading{Cascade decomposition}
The same cascade structure that creates the MNAR bias also provides an exact correction.
For a depth-1 path, the observed mean is already an unbiased estimate:
\[
\hat\mu(\ell_1)=\bar A_{\ell_1}.
\]
For a depth-2 path $(\ell_1,\ell_2)$, success occurs if $\ell_1$ succeeds, or if $\ell_1$ fails and $\ell_2$ then succeeds.
These events are disjoint, so
\begin{equation}
  \mu(\ell_1,\ell_2)
  =
  \mu(\ell_1)
  +
  \bigl(1-\mu(\ell_1)\bigr)\cdot
  \Pr[\ell_2\text{ succeeds} \mid \ell_1\text{ fails}].
  \label{eq:decomp}
\end{equation}
The conditional term is exactly what cascade sampling observes for column $(\ell_1,\ell_2)$.
Hence, if $\bar A_{(\ell_1,\ell_2)}$ denotes the raw sample mean of that column, then
\[
  \hat\mu(\ell_1,\ell_2)
  =
  \hat\mu(\ell_1)
  +
  \bigl(1-\hat\mu(\ell_1)\bigr)\cdot
  \bar A_{(\ell_1,\ell_2)}.
\]
The same recursion applies at depth 3 and beyond: each observed column mean estimates the conditional success probability at that depth given that all earlier stages failed.
The estimator is therefore consistent for every path $p$, and the MNAR pattern that breaks direct averaging becomes, under this decomposition, exactly the right conditioning.

\heading{Low-rank smoothing}
At a 5\% profiling budget, depth-3 columns receive only $\approx\!20$--$80$ observations each, making the cascade decomposition estimates noisy for deep paths.
As a practical variance-reduction step, \sys applies a rank-1 SVD projection to the depth-3 conditional accuracy estimates before substituting them into the decomposition.

\heading{End-to-end effectiveness}
These techniques are complementary: checkpointing reduces redundant execution, cascade sampling and subtree fill-in maximize useful observations, and cascade decomposition corrects the MNAR bias that defeats direct averaging and vanilla matrix completion.
As we show in \S\ref{sec:evaluation:matrix}, together they reduce path-level MAE to roughly 1\% using only 2\% of the cost of exhaustive profiling, whereas standard matrix-completion baselines remain at 5\% error.
This makes sparse offline trie annotation accurate enough to support online model selection.

\subsection{Online Model Selection}
\label{sec:design:dynamic}
At runtime, \sys\ decides per configurable stage.
After each stage finishes, the controller observes the realized execution prefix $u$ together with the cumulative latency so far $T_u$.
It then re-roots the annotated trie at $u$, searches only the descendants of $u$, executes the next stage/model action on the best feasible suffix, updates $(u, T_u)$, and repeats.
Thus, \sys\ never commits to a fixed workflow-level plan at admission time; it interleaves execution and control, allowing later choices to depend on the actual progress of the request.

This receding-horizon design is useful whenever realized behavior differs from offline averages~\cite{constrainedmodelpredictivecontrol}.
If an early stage is slower than expected, the remaining budget shrinks and the controller can switch to a faster suffix.
If earlier stages finish quickly, \sys\ can spend the saved budget on a stronger downstream model or an additional refinement step.
The same mechanism also handles loops naturally: each loop iteration moves the request to a deeper prefix in the trie, and \sys\ solves the same optimization problem again on the remaining subtree.

Because expected accuracy, cost, and latency are monotone along every root-to-leaf path, search can prune large parts of the trie.
For objectives that minimize cost subject to an accuracy floor, we use pruned DFS.
Once a node already satisfies the accuracy target, exploring its descendants cannot improve that branch, since all descendants have weakly higher cost and latency.
After the first feasible node is found, its objective value becomes an incumbent bound, so any prefix whose current cost or latency already exceeds that bound can be discarded.
For objectives that maximize accuracy under cost or latency caps, pruning is weaker: internal-node accuracy does not justify pruning, because descendants may still improve accuracy while remaining feasible.
In that case, pruning is driven only by prefixes that already violate the relevant budgets.

\heading{Runtime budget updates}
These pruning rules are re-applied after every stage invocation using the realized execution so far.
If the current prefix is $u$, the controller optimizes over descendants $v \succeq u$ using the remaining budgets, equivalently the incremental estimates
\[
\Delta \hat T_u(v) = \hat{\bar T}(v) - \hat{\bar T}(u).
\]
The trie's accuracy and cost estimates do not change during execution; what changes is which suffixes remain feasible after accounting for realized latency.
Dynamic adaptation is therefore not a different optimization problem, but the same trie search repeated as new execution feedback arrives.

\heading{Load-aware latency adjustment}
If \sys\ has a current estimate of queueing delay or service slowdown for each serving engine $e$, it can incorporate that signal directly into search.
For any suffix $v \succeq u$, \sys\ replaces the offline latency estimate with
\[
\Delta \hat T^{\,\mathrm{live}}_u(v)
=
\Delta \hat T_u(v)
+
\sum_{e \in \mathrm{engines}(v \setminus u)} \delta_e(t),
\]
where $\delta_e(t)$ is the current expected delay for engine $e$.
This inflates paths that rely on congested backends and steers the controller toward suffixes less likely to violate the request's latency target.

\subsection{Implementation}
\label{sec:design:impl}

\sys is implemented as a control layer on top of an existing workflow runtime, LangGraph~\cite{langgraph}.
The runtime maintains a typed workflow state that stores the request context, intermediate artifacts, retry counters, the realized execution prefix, and per-stage metadata.
\sys wraps each configurable LLM stage with a routing layer that selects the model or endpoint for the next invocation, executes it, and records the resulting measurements.
Fixed tool stages (e.g., SQL execution) are executed by the underlying workflow and return structured feedback that determines the next control-flow transition.

The prototype supports multiple serving backends, including Amazon Bedrock~\cite{aws_bedrock}, Gemini, OpenAI-based backends, and SGLang~\cite{zheng2023sglang}.
Each invocation logs the selected model or endpoint together with token usage and latency statistics.
For self-hosted backends such as SGLang, \sys records time to first token, decode time, and token counts; for API-hosted backends such as Bedrock, it records provider-reported latency together with prompt and output tokens.
These traces are used both to reconstruct offline cost and latency annotations and to update remaining budgets online during execution.

The execution trie is materialized through checkpointed workflow prefixes rather than by replaying every path from the root.
The Profiler expands the search space depth by depth and runs each request--prefix pair in an isolated subprocess, which localizes failures and produces one structured execution record per node.
Deeper workers resume from serialized parent checkpoints, execute only the remaining suffix, and emit updated checkpoints containing prefix metadata, evaluation results, and termination status.
The Trie Estimator merges checkpoints by request and prefix, reconstructs path-level cost and latency, aggregates node statistics, applies subtree fill-in and the sparse estimator from \S\ref{sec:design:build}, and produces the annotated trie used by the online \sys Controller.

\subsection{Discussion}
\label{sec:design:discussion}

\heading{Context reuse}
Switching models across refinement iterations can reduce context reuse: a later step served by a different model or endpoint may require re-sending the accumulated context and paying the prefill cost again~\cite{zheng2023sglang,pan2025kvflow,autellix,helium,llmcdn,lmcache,cacheblend,cachegen}.
In our workloads this overhead was modest, and \sys naturally accounts for it because profiling is performed over full workflow paths, so any extra latency or cost from re-prefill is already reflected in the trie annotations.

\heading{Non-LLM stages}
Tool stages such as retrieval, SQL execution, or external API calls are treated as fixed transitions in the workflow template. They do not introduce branching in the trie, but their measured latency and cost are folded into cumulative path metrics, allowing \sys to reason jointly about LLM choices and tool overheads.

\heading{Distribution mismatch}
The trie also serves as a monitoring abstraction: \sys can compare live path statistics against offline annotations and detect when observed latency or success rates drift away from the profiling distribution~\cite{mallick2022matchmaker,lewis2022augur}. When that happens, the right response is to refresh or recalibrate the trie using newer requests.

\heading{Resource allocation and hardware choices}
Resource provisioning and hardware allocation are complementary to \sys's per-invocation control. A higher-level planner can choose replicas, hardware, or endpoints~\cite{crankshaw2020inferline,romero2021llama}, while \sys selects among those options online; the trie can be extended so that each action is a model--endpoint pair rather than just a model. We leave this joint, multi-timescale optimization to future work.

\section{Evaluation}
\label{sec:evaluation}

We evaluate \sys along the following questions: (i) How much does finer-grain control improve over Murakkab-style workflow-level control in refinement loops? (ii) How closely can sparse profiling recover the decisions of a fully profiled trie and at what cost? (iii) How well does dynamic per-step control handle latency SLOs under runtime variance and load?

\subsection{Experimental Setup}
\label{sec:evaluation:setup}

\heading{Workflows}
We evaluate \sys on two refinement-heavy workloads. The first workload is an NL2SQL pipeline based on a long-context NL2SQL system~\cite{chung2025longcontext}. We use two variants of this workflow. \mbox{NL2SQL-8} exposes eight candidate models and reaches total depth~3, consisting of one generation step and up to two refinement steps. \mbox{NL2SQL-2} exposes two candidate models and reaches total depth~4, consisting of one generation step and up to three refinement steps.
The second workload is a self-reflection math question-answering workflow~\cite{renze2024selfreflection}; we refer to this workload as MathQA.
MathQA is a single-stage reflection workflow with up to six invocations of the same logical stage and four candidate models.
These depths define the maximum horizon we profile, but the loop budget used for each request is chosen based on the target SLOs (accuracy, cost or latency) and the offline estimation of the managers rather than fixed arbitrarily or by workflow-specific stop semantics.

\heading{Models}
For \mbox{NL2SQL-8}, we use Gemma-3-27B~\cite{gemma327b}, Sonnet-4.6~\cite{sonnet46}, Kimi-K2.5~\cite{kimi}, Qwen3-32B~\cite{qwen}, GLM-4.7~\cite{glm47}, Llama-3.3-70B~\cite{llama33}, DeepSeek-V3.2~\cite{deepseek}, and gpt-oss-120b~\cite{gptoss120b}. For \mbox{NL2SQL-2}, we use Gemma-3-27B and Sonnet-4.6. MathQA uses Gemma-3-27B, Sonnet-4.6, Kimi-K2.5, and Qwen3-32B.

\heading{Baselines}
Our main baseline is the workflow-level control space of Murakkab~\cite{chaudhry2025murakkab}. Murakkab can choose different models for different stage templates and can also choose the maximum number of refinement iterations, but it cannot choose a different model for different invocations of the same stage within one request. In NL2SQL, Murakkab can choose one model for generation, one model for refinement, and a refinement-depth cap. In MathQA, because the workflow has only one repeated stage, Murakkab can choose only one model for the entire workflow. Unless stated otherwise, Murakkab and \sys\ (full) use full offline profiling. \sys\ (sparse) uses only 2\% of the full offline LLM profiling cost with the trie-estimation methods of \S\ref{sec:design:build}.

\heading{Execution environment}
We serve all workflow models through Amazon Bedrock~\cite{aws_bedrock}, and we use Amazon EC2~\cite{aws_ec2} for the client machines. The client machine for NL2SQL is an \texttt{m6a.24xlarge} instance and the client machine for MathQA is an \texttt{m6a.8xlarge} instance.


\subsection{Gain over Murakkab's Workflow-Level Control}
\label{sec:evaluation:slo}

\begin{figure*}[t]
    \centering
    \includegraphics[width=0.95\linewidth]{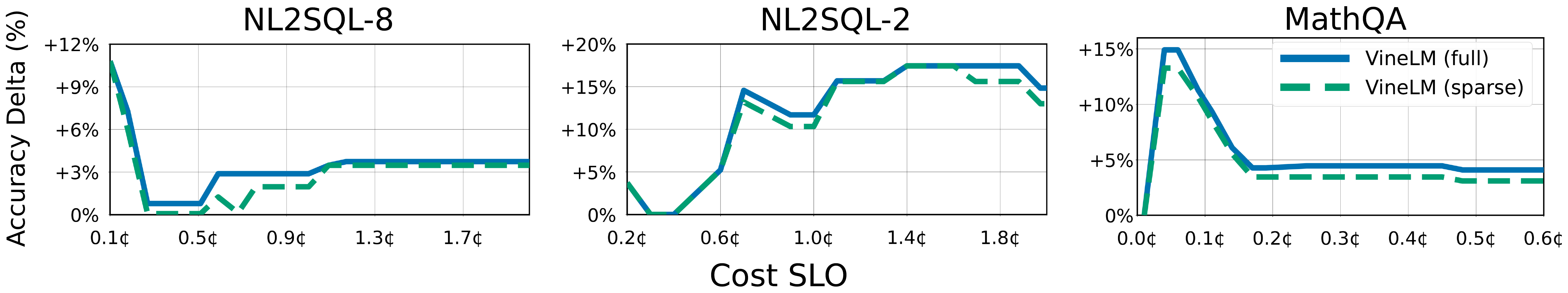}
    \caption{Accuracy delta over Murakkab for three workflows: \mbox{NL2SQL-8}, \mbox{NL2SQL-2}, and MathQA.}
    \label{fig:evaluation:full-tree:accuracy_delta_triptych}
\end{figure*}

We compare \sys\ to the best Murakkab-style workflow-level configuration under the same cost SLO.
Figure~\ref{fig:evaluation:full-tree:accuracy_delta_triptych} reports the accuracy gain of \sys\ over Murakkab on \mbox{NL2SQL-8}, \mbox{NL2SQL-2}, and MathQA.
The solid line uses the fully profiled trie; the dashed line uses sparse profiling at 0.2-2\% of the full profiling cost.
Across all three workloads, \sys\ consistently outperforms the best static workflow-level choice, and sparse \sys\ preserves most of that gain despite using only a tiny fraction of the profiling cost.

\sys\ optimizes over a richer action space.
For the case of \mbox{NL2SQL-8}, this is 136 Murakkab configurations versus 584 trie paths; for \mbox{NL2SQL-2}, it is 14 versus 30.
Even in \mbox{NL2SQL-8}, where Murakkab already has a relatively rich workflow-level space, \sys\ still delivers a consistent positive accuracy delta across the cost range.
This is the hardest setting for \sys\ to separate from Murakkab, because Murakkab already has many static configurations to choose from.
The gain is larger in \mbox{NL2SQL-2}.
With fewer models, Murakkab has even less ability to approximate the right repair sequence, while \sys\ can still adapt each repair invocation separately.
As a result, the advantage of modeling the loop as a sequence of prefix-conditioned decisions becomes more pronounced.
Sparse \sys\ retains most of that advantage.

MathQA shows a smaller but still positive delta for a different reason.
Because MathQA is a repeated-reflection workflow, Murakkab effectively commits to one model for the reflection process, whereas \sys\ can mix models across rounds.
The gain is smaller not because fine-grained control stops helping, but because baseline accuracy is already higher, leaving less headroom for any controller to improve end-to-end success.

The gain is not just that \sys\ has more choices.
Murakkab optimizes over workflow-level profiles: one generation model, one repair model, and a loop horizon or round count, fixed before the request begins.
That view collapses the loop into a single average plan.
In particular, when estimating the value of later repair rounds, it cannot condition on which requests actually survive to those rounds.
Requests that would already have finished after Generation or after the first Repair are still mixed into the same workflow-level expectation.
\sys\ does not have this problem.
Because the execution trie is indexed by prefixes and the cascade estimator recovers prefix-conditioned success probabilities, \sys\ prices a continuation only on the subset of requests that actually reaches that prefix.
This lets \sys\ value later repair rounds more accurately and decide whether another step is still worth the remaining budget.

\subsection{A Deep Dive into Trie-Filling Methods}
\label{sec:evaluation:matrix}

We next ask how closely \sys\ can recover the fully profiled frontier after observing only a small fraction of request--path executions. Coverage denotes the fraction of the full offline LLM profiling cost spent on sparse sampling. The goal in this subsection is not to reconstruct every missing entry of $A$, but to predict the column means that determine path selection.

We compare six estimators. First, is \emph{direct average}: for each path, predict its column mean by averaging the observed entries in that column.
Under cascade sampling, this estimator is badly biased for deeper paths because those paths are only observed on harder requests.
Second, \emph{prefix fill-in + avg}: first apply prefix-success closure, then average the observed entries in each column.
Prefix fill-in marks all descendants of an observed success as successful, because once a prefix succeeds, every extension of that prefix also succeeds.
Third, \emph{prefix fill-in + hard impute}: after prefix fill-in, apply a low-rank matrix-completion step to impute the remaining unobserved cells, then average the completed columns.
Fourth, \emph{prefix fill-in + XGBoost}: after prefix fill-in, apply XGBoost~\cite{xgboost} which predicts each column mean from hand-designed path and observation features, including path depth, row and column observation counts, row and column mean success rates, prefix values and prefix means at several depths, sibling statistics, and model power-score summaries along the path.
Fifth, \emph{\sys-Lite}: estimate path means through the cascade decomposition of \S\ref{sec:design:build}, which corrects the MNAR bias by treating deeper observations as conditional accuracies.
Sixth, \emph{\sys{}}: adds the low-rank smoothing step on top of conditional decomposition. 

\begin{figure}[t]
    \centering
    \includegraphics[width=0.9\linewidth]{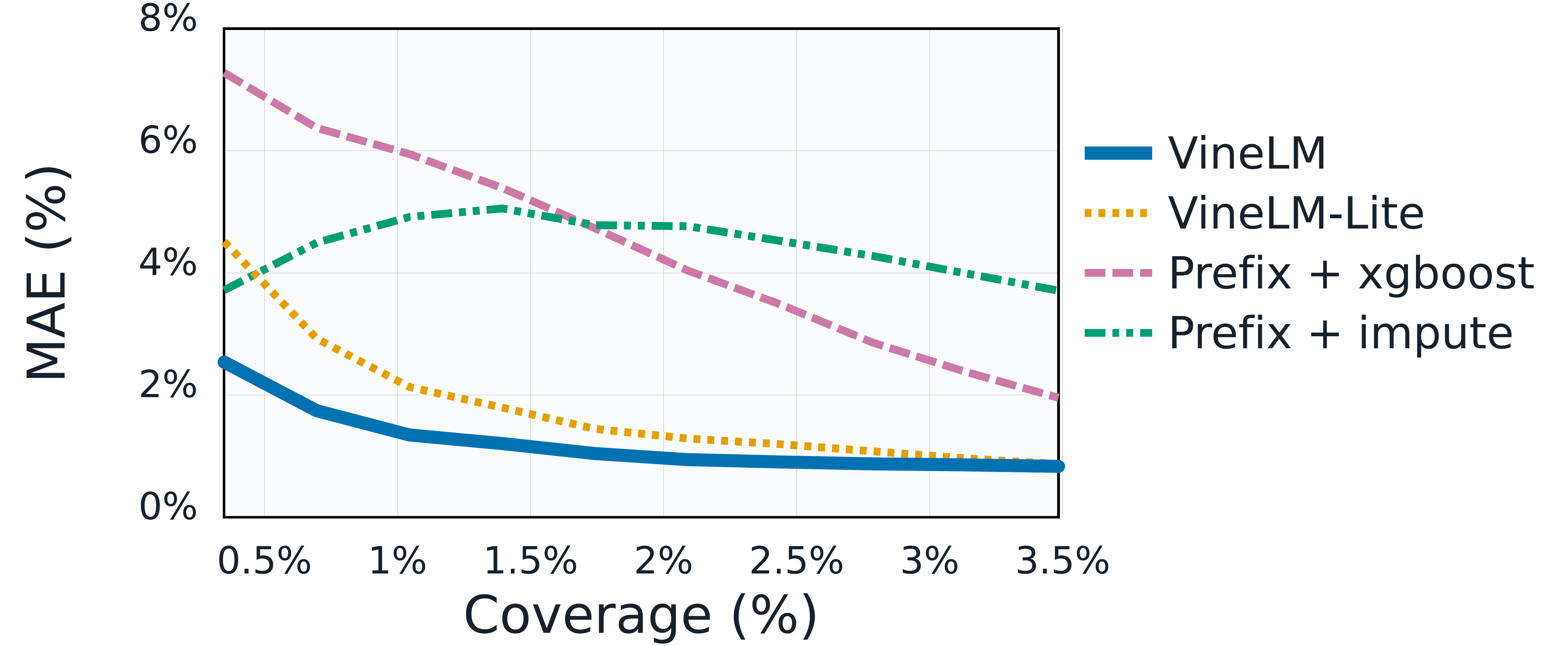}
    \caption{Column-mean prediction error versus profiling coverage. Coverage is the fraction of full offline LLM profiling cost used for sparse sampling.}
    \label{fig:evaluation:cheap-profiling:average_column_error:mae_graph}
\end{figure}

Figure~\ref{fig:evaluation:cheap-profiling:average_column_error:mae_graph} plots column-mean MAE as a function of coverage. \sys\ attains the lowest error across the full range and reaches about 1\% MAE at 2\% coverage. The ordering matches the MNAR analysis in \S\ref{sec:formulation:challenges}: methods that ignore the depth-dependent sampling bias misestimate the deeper columns that determine the high-accuracy end of the frontier.

\begin{table}[t]
\centering
\caption{Column-level error summary at 2\% cost coverage for NL2SQL. Signed error is prediction minus ground-truth column mean.}
\begin{tabular}{lrrr}
\hline
Method & Mean Signed & Mean Abs. & Max Abs. \\
\hline
average & -49.53\% & 49.53\% & 76.46\% \\
prefix + avg & 20.50\% & 20.50\% & 33.17\% \\
prefix + impute & 4.78\% & 4.78\% & 12.98\% \\
prefix + xgboost & 3.08\% & 4.73\% & 15.57\% \\
\sys-Lite & \textbf{0.04\%} & 1.45\% & 6.29\% \\
\sys & 0.07\% & \textbf{1.04\%} & \textbf{4.33\%} \\
\hline
\end{tabular}
\label{tab:tianyi-5pct-column-error-summary}
\vspace{-0.25in}
\end{table}

To analyze the methods more closely, we fix coverage at 2\% and examine the full error distribution. Table~\ref{tab:tianyi-5pct-column-error-summary} shows both the scale and the direction of the mispredictions. Signed error separates pessimism from optimism. The direct average is strongly pessimistic because direct averaging observes only the suffix-conditioned hard subpopulation. Prefix-based methods are optimistic on average.
\sys variants are nearly unbiased in mean signed error, which shows that conditional decomposition removes the main source of bias.
The remaining difference between \sys and \sys-Lite is mostly in variance and tail behavior: \sys reduces mean absolute error from 0.0145 to 0.0104 and max absolute error from 0.0629 to 0.0433.
This difference matters because one large column misprediction can change the selected path even when average error remains small.

\begin{figure}[t]
    \centering
    \includegraphics[width=0.9\linewidth]{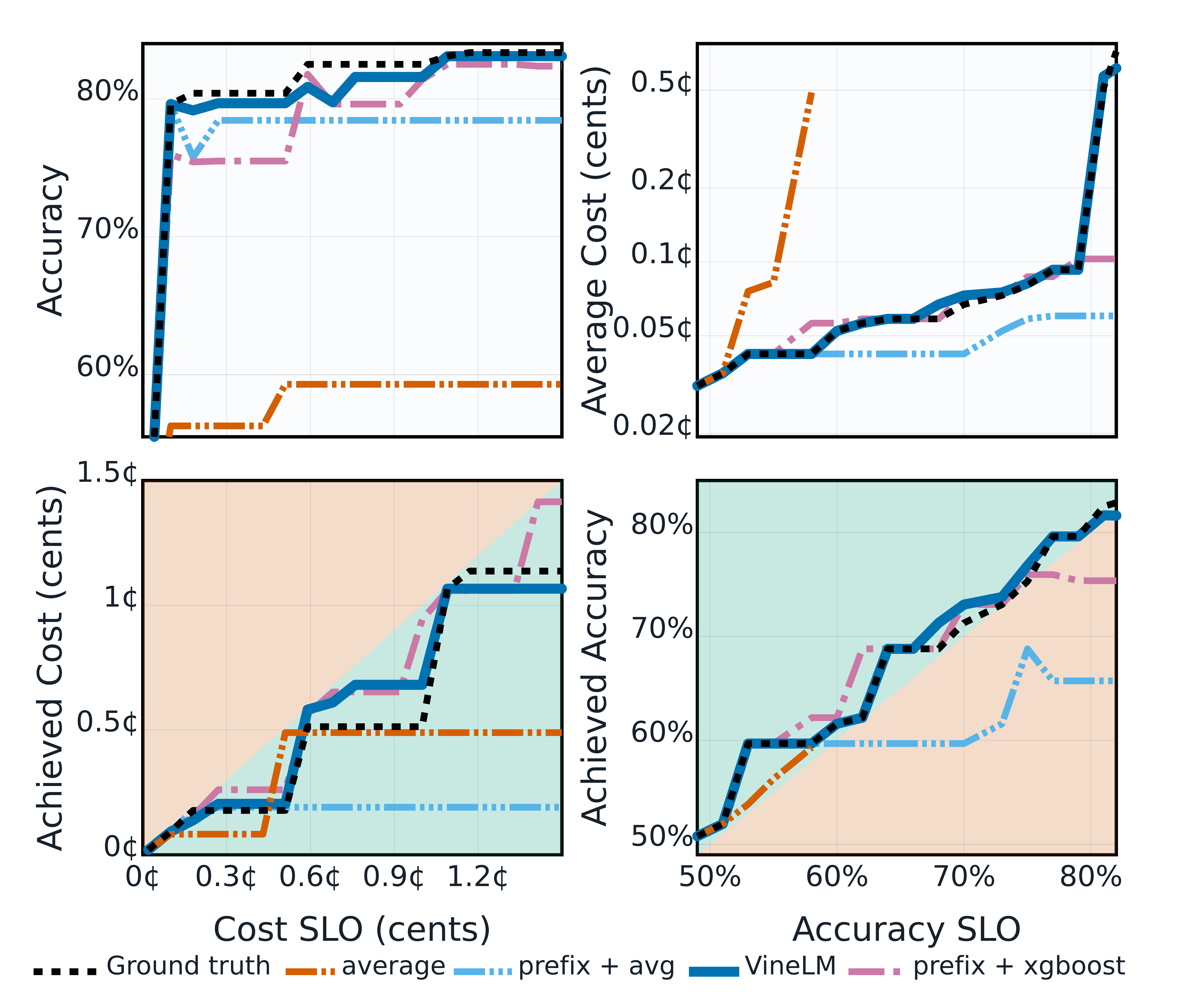}
    \caption{Maximum accuracy under a cost SLO (left) and minimum average cost under an accuracy SLO (right), at 2\% profiling coverage.}
    \label{fig:evaluation:cheap-profiling:imputation-policies:max_accuracy_given_cost}
    \vspace{-.25in}
\end{figure}

We then keep coverage fixed at 2\% and rerun the policy search with predicted column means. Figure~\ref{fig:evaluation:cheap-profiling:imputation-policies:max_accuracy_given_cost} compares the resulting maximum-accuracy-under-cost curves against the fully profiled ground truth. \sys\ tracks the ground-truth line closely in both the achieved-accuracy panel and the achieved-cost panel, which shows that low column-mean error translates into high-fidelity policy selection.

The same figure shows the comparison for the minimum-cost-under-accuracy objective. \sys\ again follows the ground-truth frontier closely. Some baselines appear cheaper in the top panel, especially prefix fill-in + avg, but the bottom panel shows why: those methods fall below the $y=x$ line and therefore violate the accuracy SLO. The same failure mode appears for XGBoost at higher accuracy targets.
These methods do not adequately correct the sampling bias in \S\ref{sec:formulation:challenges}, and they also underprofile the deeper paths that are usually necessary to reach higher accuracy.
The direct-average baseline fails even earlier. Its curve terminates around 57\% accuracy because it severely mispredicts the deep paths that dominate the high-accuracy regime.

\begin{table}[t]
\caption{Profiling cost in dollars.}
\begin{tabular}{lrrrr}
\hline
Workflow & \sys & Chkpt & Full & Ratio \\
\hline
MathQA-4 & 28.16 & 563.18 & 15073.67 & 535.29x \\
NL2SQL-2 & 5.28 & 105.56 & 250.49 & 47.44x \\
NL2SQL-8 & 48.19 & 963.89 & 2764.21 & 57.36x \\

\hline
\end{tabular}

\label{tab:profiling-cost}
\end{table}

\heading{Profiling Cost}
Table~\ref{tab:profiling-cost} reports three profiling regimes: \sys's sparse profiler (\sys), full checkpointed exhaustive profiling (Chkpt), and naive exhaustive profiling from the root on every leaf path (Full).
Even without sparsity, checkpointing substantially reduces full-profiling cost by reusing shared prefixes: compared with the naive Full baseline, checkpointing alone reduces cost by 2.37$\times$ on \mbox{NL2SQL-2}, 2.87$\times$ on \mbox{NL2SQL-8}, and 26.77$\times$ on MathQA-4.
Combining checkpointing with sparse profiling yields much larger savings.
Relative to Full, \sys reduces profiling cost by 47.44$\times$ on \mbox{NL2SQL-2}, 57.36$\times$ on \mbox{NL2SQL-8}, and 535.29$\times$ on MathQA-4.
These results show that shared-prefix reuse is already important for exhaustive profiling, and that \sys's full sparse-trie pipeline makes offline trie annotation cheap enough to be practical.
As expected, gains are larger in complex, deep workflows with many available models to choose from.

\subsection{Latency SLO and Online Model Selection}
\label{sec:evaluation:latency}
So far, we have focused on cost and accuracy. Latency is different because stage-time variability can create per-request SLO violations even when the offline plan is optimal in expectation. We therefore compare three policies: (i) Murakkab, which commits to a single full path at admission time, (ii) a dynamic load-unaware policy that recomputes the best remaining suffix after each stage using realized elapsed time, and (iii) a dynamic load-aware policy that also inflates latency estimates using the current load model.
All three policies optimize for accuracy subject to a latency SLO.

The load model comes from a separate queueing experiment on SGLang setup. All requests were real NL2SQL problems. We used one fixed problem as the target request and injected artificial load by sending $N$ higher-priority dummy requests in parallel, with $N \in \{0,1,2,4,8,16,32\}$. Dummy requests were drawn randomly from the full NL2SQL set. We delayed the target request by 2000\,ms so that most dummy requests were already queued, repeated each load level 50 times, and flushed the KV cache between runs. We then fit the utilization-conditioned slowdown curve used to inflate recorded latencies during evaluation.

\begin{figure*}[t]
    \centering
    \includegraphics[width=1\linewidth]{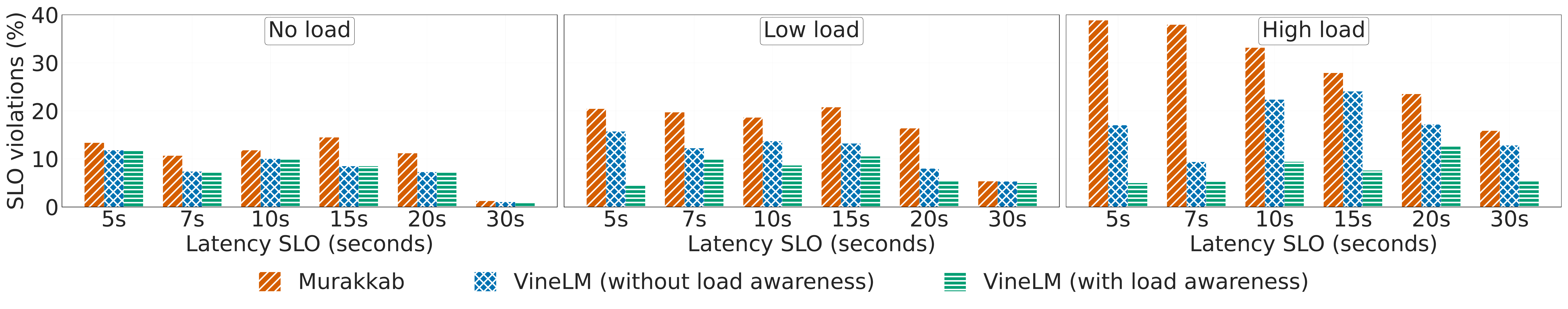}
    \caption{Latency-SLO violation rate for Murakkab, dynamic load-unaware control, and dynamic load-aware control. Dynamic replanning reduces violations, and load-aware replanning reduces them further under load.}
    \label{fig:evaluation:slack-based:dynamic}
\end{figure*}

Figure~\ref{fig:evaluation:slack-based:dynamic} shows that per-stage replanning reduces latency-SLO violations relative to Murakkab across the tested SLO range. Under additional load, the load-aware policy reduces violations further by steering requests away from more overloaded, slower paths.
Violation counts do not necessarily decrease as the latency SLO increases.
A looser latency budget lets the controller choose slower and more accurate paths, so the operating point remains close to the latency frontier.
Residual latency misprediction can therefore still produce violations even at larger SLOs.

\begin{table}[t]
\caption{Overhead of the dynamic method. Overhead (\%) shows the overhead of the dynamic method compared to the average latency for the fastest LLM call in the workflow.}
\begin{tabular}{lrrr}
\toprule
Workflow & Mean ($\mu$s) & Overhead (\%) \\
\midrule
  MathQA-4 & 6921.2 & 0.592 \\
  NL2SQL-2 & 33.0 & 0.001 \\
  NL2SQL-8 & 1312.0 & 0.052 \\
\bottomrule
\label{tab:evaluation:slack-based:online-method-overhead}
\end{tabular}
\vspace{-0.25in}
\end{table}



Online selection adds a small search cost after each stage because it re-roots the trie and recomputes the best remaining suffix. Table~\ref{tab:evaluation:slack-based:online-method-overhead} shows that this cost is negligible relative to LLM calls.
Even the largest configuration we measure---the depth-6 MathQA workflow with four models---adds 6.921\,ms per replanning step, which is only 0.592\% of the fastest LLM call in that workflow.
The smaller configurations are much cheaper: 33\,$\mu$s for NL2SQL with two models and 1.312\,ms for NL2SQL with eight models. As expected, the overhead increases with both workflow depth and the number of candidate models, but it remains below 1\% in all measured cases.

\section{Related Work}
\label{sec:related}

\noindent\textbf{Agentic workflow serving systems.}
The system closest to \sys is Murakkab~\cite{chaudhry2025murakkab}.
Murakkab decouples logical workflow specification from execution choices and uses offline profiles, an SLO-aware optimizer, and an adaptive runtime to choose among pre-profiled workflow-level configurations.
Our main distinction is decision granularity.
Murakkab commits each request to a workflow-level plan that binds one model to each configurable LLM stage and fixes loop-related knobs up front.
By contrast, \sys controls a workflow at the level of stage invocations.

\noindent\textbf{LLM routing and cascades.}
A large recent literature studies how to route a query to an appropriate LLM to improve the cost-quality tradeoff.
RouteLLM~\cite{ong2025routellm}, Hybrid LLM~\cite{ding2024hybridllm}, IRT-Router~\cite{irtrouter}, and Causal LLM Routing~\cite{tsiourvas2025causal} learn prompt-level routing policies using preference data, predicted difficulty, or observational logs, while FrugalGPT~\cite{chen2024frugalgpt} and later work on routing--cascading unification~\cite{dekoninck2025unified} study sequential escalation from cheaper to stronger models.
These methods solve a flatter problem than ours: choose one model, or a short cascade, for a single query.
\sys instead controls multi-stage workflows with tools and loops, where an early decision changes not only immediate cost and latency but also the remaining budget and whether later stages are entered at all.
However, similar signals such as question complexity could be used to bias \sys's initial search toward cheaper or stronger prefixes, or to maintain separate trie annotations for different difficulty buckets.

\noindent\textbf{Classical ML serving and pipeline management.}
Classical ML-serving systems address neighboring problems, but under different abstractions.
Clipper~\cite{crankshaw2017clipper} introduced online model selection and ensembling over a flat set of models.
INFaaS~\cite{romero2021infaas} supports per-request performance and accuracy targets, but still operates over single-model variants and model/hardware profiles.
InferLine~\cite{crankshaw2020inferline} and Llama~\cite{romero2021llama} move closer to workflow-level optimization by planning and adapting multi-stage pipelines under latency objectives.
These systems demonstrate the value of workflow-aware planning and online adaptation, but they target classical pipelines where stage choices primarily affect latency, throughput, and resource efficiency.
\sys instead targets agentic LLM workflows with loops, where invocation-level model choices also affect semantic accuracy and future control flow.

\section{Conclusion}
\label{sec:conclusion}

\sys uses an annotated execution trie to make fine-grained control practical for agentic workflows.
Hence, its controller can avoid the common pitfall of coarse workflow-level policies that bind one model to an entire refinement loop and cannot revise downstream choices as execution unfolds.
By combining checkpointing, cascade profiling, and rerooted online control, \sys improves the cost--latency--accuracy frontier for refinement-heavy workloads while keeping profiling overhead low.
Across NL2SQL and math reasoning workflows, \sys achieves up to 18\% higher accuracy at the same budget, reduces offline profiling cost by 98\%, and cuts latency-SLO violations by up to 85\% relative to coarse workflow-level baselines.
These results suggest that agentic workflows should be controlled with online policies over execution prefixes, rather than with fixed workflow configurations.

\normalem

\bibliographystyle{plain}
\bibliography{main}

@misc{aws_bedrock,
  author       = {{Amazon Web Services}},
  title        = {{Amazon Bedrock}},
  year         = {2026},
  url          = {https://aws.amazon.com/bedrock/},
  note         = {Accessed: 2026-04-02}
}

@misc{aws_ec2,
  author       = {{Amazon Web Services}},
  title        = {{Amazon Elastic Compute Cloud (Amazon EC2)}},
  year         = {2026},
  url          = {https://aws.amazon.com/ec2/},
  note         = {Accessed: 2026-04-02}
}

@misc{gemma327b,
  author       = {{Google}},
  title        = {{gemma-3-27b-it Model Card}},
  year         = {2025},
  publisher    = {Hugging Face},
  url          = {https://huggingface.co/google/gemma-3-27b-it},
  note         = {Accessed: 2026-04-02}
}

@misc{sonnet46,
  author       = {{Anthropic}},
  title        = {{Claude Sonnet}},
  year         = {2024},
  url          = {https://www.anthropic.com/claude/sonnet},
  note         = {Accessed: 2026-04-02}
}

@misc{glm47,
  author       = {{Zhipu AI}},
  title        = {{GLM-4.7 Technical Blog}},
  year         = {2025},
  url          = {https://z.ai/blog/glm-4.7},
  note         = {Accessed: 2026-04-02}
}

@misc{llama33,
  author       = {{Meta}},
  title        = {{Llama-3.3-70B-Instruct Model Card}},
  year         = {2024},
  publisher    = {Hugging Face},
  url          = {https://huggingface.co/meta-llama/Llama-3.3-70B-Instruct},
  note         = {Accessed: 2026-04-02}
}

@misc{gptoss120b,
  author       = {{OpenAI}},
  title        = {{gpt-oss-120b \& gpt-oss-20b Model Card}},
  year         = {2025},
  eprint       = {2508.10925},
  archivePrefix= {arXiv},
  primaryClass = {cs.CL},
  publisher    = {Hugging Face},
  url          = {https://huggingface.co/openai/gpt-oss-120b},
  note         = {Accessed: 2026-04-02}
}

@misc{deepseek,
      title={DeepSeek-V3.2: Pushing the Frontier of Open Large Language Models}, 
      author={DeepSeek-AI},
      year={2025},
      eprint={2512.02556},
      archivePrefix={arXiv},
      primaryClass={cs.CL},
      url={https://arxiv.org/abs/2512.02556}, 
}

@misc{qwen,
      title={Qwen3 Technical Report}, 
      author={An Yang and Anfeng Li and Baosong Yang and Beichen Zhang and Binyuan Hui and Bo Zheng and Bowen Yu and Chang Gao and Chengen Huang and Chenxu Lv and Chujie Zheng and Dayiheng Liu and Fan Zhou and Fei Huang and Feng Hu and Hao Ge and Haoran Wei and Huan Lin and Jialong Tang and Jian Yang and Jianhong Tu and Jianwei Zhang and Jianxin Yang and Jiaxi Yang and Jing Zhou and Jingren Zhou and Junyang Lin and Kai Dang and Keqin Bao and Kexin Yang and Le Yu and Lianghao Deng and Mei Li and Mingfeng Xue and Mingze Li and Pei Zhang and Peng Wang and Qin Zhu and Rui Men and Ruize Gao and Shixuan Liu and Shuang Luo and Tianhao Li and Tianyi Tang and Wenbiao Yin and Xingzhang Ren and Xinyu Wang and Xinyu Zhang and Xuancheng Ren and Yang Fan and Yang Su and Yichang Zhang and Yinger Zhang and Yu Wan and Yuqiong Liu and Zekun Wang and Zeyu Cui and Zhenru Zhang and Zhipeng Zhou and Zihan Qiu},
      year={2025},
      eprint={2505.09388},
      archivePrefix={arXiv},
      primaryClass={cs.CL},
      url={https://arxiv.org/abs/2505.09388}, 
}

@misc{kimi,
      title={Kimi K2.5: Visual Agentic Intelligence}, 
      author={Kimi Team},
      year={2026},
      eprint={2602.02276},
      archivePrefix={arXiv},
      primaryClass={cs.CL},
      url={https://arxiv.org/abs/2602.02276}, 
}

@article{constrainedmodelpredictivecontrol,
author = {Mayne, D. Q. and Rawlings, J. B. and Rao, C. V. and Scokaert, P. O. M.},
title = {Survey Constrained model predictive control: Stability and optimality},
year = {2000},
issue_date = {June, 2000},
publisher = {Pergamon Press, Inc.},
address = {USA},
volume = {36},
number = {6},
issn = {0005-1098},
url = {https://doi.org/10.1016/S0005-1098(99)00214-9},
doi = {10.1016/S0005-1098(99)00214-9},
journal = {Automatica},
month = jun,
pages = {789–814},
numpages = {26}
}

@article{tailatscale,
author = {Delimitrou, Christina and Kozyrakis, Christos},
title = {Amdahl's law for tail latency},
year = {2018},
issue_date = {August 2018},
publisher = {Association for Computing Machinery},
address = {New York, NY, USA},
volume = {61},
number = {8},
issn = {0001-0782},
url = {https://doi.org/10.1145/3232559},
doi = {10.1145/3232559},
journal = {Commun. ACM},
month = jul,
pages = {65–72},
numpages = {8}
}

@article{amdahltail,
author = {Delimitrou, Christina and Kozyrakis, Christos},
title = {Amdahl's law for tail latency},
year = {2018},
issue_date = {August 2018},
publisher = {Association for Computing Machinery},
address = {New York, NY, USA},
volume = {61},
number = {8},
issn = {0001-0782},
url = {https://doi.org/10.1145/3232559},
doi = {10.1145/3232559},
journal = {Commun. ACM},
month = jul,
pages = {65–72},
numpages = {8}
}

@misc{react,
      title={ReAct: Synergizing Reasoning and Acting in Language Models}, 
      author={Shunyu Yao and Jeffrey Zhao and Dian Yu and Nan Du and Izhak Shafran and Karthik Narasimhan and Yuan Cao},
      year={2023},
      eprint={2210.03629},
      archivePrefix={arXiv},
      primaryClass={cs.CL},
      url={https://arxiv.org/abs/2210.03629}, 
}

@misc{llmcdn,
      title={Do Large Language Models Need a Content Delivery Network?}, 
      author={Yihua Cheng and Kuntai Du and Jiayi Yao and Junchen Jiang},
      year={2024},
      eprint={2409.13761},
      archivePrefix={arXiv},
      primaryClass={cs.CL},
      url={https://arxiv.org/abs/2409.13761}, 
}

@inproceedings{cachegen,
author = {Liu, Yuhan and Li, Hanchen and Cheng, Yihua and Ray, Siddhant and Huang, Yuyang and Zhang, Qizheng and Du, Kuntai and Yao, Jiayi and Lu, Shan and Ananthanarayanan, Ganesh and Maire, Michael and Hoffmann, Henry and Holtzman, Ari and Jiang, Junchen},
title = {CacheGen: KV Cache Compression and Streaming for Fast Large Language Model Serving},
year = {2024},
isbn = {9798400706141},
publisher = {Association for Computing Machinery},
address = {New York, NY, USA},
url = {https://doi.org/10.1145/3651890.3672274},
doi = {10.1145/3651890.3672274},
booktitle = {Proceedings of the ACM SIGCOMM 2024 Conference},
pages = {38–56},
numpages = {19},
location = {Sydney, NSW, Australia},
series = {ACM SIGCOMM '24}
}

@inproceedings{cacheblend,
author = {Yao, Jiayi and Li, Hanchen and Liu, Yuhan and Ray, Siddhant and Cheng, Yihua and Zhang, Qizheng and Du, Kuntai and Lu, Shan and Jiang, Junchen},
title = {CacheBlend: Fast Large Language Model Serving for RAG with Cached Knowledge Fusion},
year = {2025},
isbn = {9798400711961},
publisher = {Association for Computing Machinery},
address = {New York, NY, USA},
url = {https://doi.org/10.1145/3689031.3696098},
doi = {10.1145/3689031.3696098},
booktitle = {Proceedings of the Twentieth European Conference on Computer Systems},
pages = {94–109},
numpages = {16},
location = {Rotterdam, Netherlands},
series = {EuroSys '25}
}

@misc{lmcache,
      title={LMCache: An Efficient KV Cache Layer for Enterprise-Scale LLM Inference}, 
      author={Yuhan Liu and Yihua Cheng and Jiayi Yao and Yuwei An and Xiaokun Chen and Shaoting Feng and Yuyang Huang and Samuel Shen and Rui Zhang and Kuntai Du and Junchen Jiang},
      year={2025},
      eprint={2510.09665},
      archivePrefix={arXiv},
      primaryClass={cs.LG},
      url={https://arxiv.org/abs/2510.09665}, 
}

@misc{softimpute,
      title={Matrix Completion and Low-Rank SVD via Fast Alternating Least Squares}, 
      author={Trevor Hastie and Rahul Mazumder and Jason Lee and Reza Zadeh},
      year={2014},
      eprint={1410.2596},
      archivePrefix={arXiv},
      primaryClass={stat.ME},
      url={https://arxiv.org/abs/1410.2596}, 
}

@article{agent_survey,
   title={A survey on large language model based autonomous agents},
   volume={18},
   ISSN={2095-2236},
   url={http://dx.doi.org/10.1007/s11704-024-40231-1},
   DOI={10.1007/s11704-024-40231-1},
   number={6},
   journal={Frontiers of Computer Science},
   publisher={Springer Science and Business Media LLC},
   author={Wang, Lei and Ma, Chen and Feng, Xueyang and Zhang, Zeyu and Yang, Hao and Zhang, Jingsen and Chen, Zhiyuan and Tang, Jiakai and Chen, Xu and Lin, Yankai and Zhao, Wayne Xin and Wei, Zhewei and Wen, Jirong},
   year={2024},
   month=mar }

@misc{crewai,
  author       = {{CrewAI}},
  title        = {CrewAI: The Leading Multi-Agent Platform},
  howpublished = {\url{https://crewai.com/}},
  year         = {2026},
  note         = {Accessed: 2026-04-02}
}

@misc{openai_agents_sdk,
  author       = {{OpenAI}},
  title        = {OpenAI Agents SDK},
  howpublished = {\url{https://openai.github.io/openai-agents-python/}},
  year         = {2026},
  note         = {Accessed: 2026-04-02}
}

@misc{strands_agents,
  author       = {{Amazon Web Services, Inc.}},
  title        = {Strands Agents — Open Source AI Agent SDK for Python \& TypeScript},
  howpublished = {\url{https://strandsagents.com/}},
  year         = {2026},
  note         = {Accessed: 2026-04-02}
}

@misc{claude_agent_sdk,
  author       = {{Anthropic}},
  title        = {Agent SDK overview - Claude API Docs},
  howpublished = {\url{https://platform.claude.com/docs/en/agent-sdk/overview}},
  year         = {2026},
  note         = {Accessed: 2026-04-02}
}

@article{lmql,
author = {Beurer-Kellner, Luca and Fischer, Marc and Vechev, Martin},
title = {Prompting Is Programming: A Query Language for Large Language Models},
year = {2023},
issue_date = {June 2023},
publisher = {Association for Computing Machinery},
address = {New York, NY, USA},
volume = {7},
number = {PLDI},
url = {https://doi.org/10.1145/3591300},
doi = {10.1145/3591300},
journal = {Proc. ACM Program. Lang.},
month = jun,
articleno = {186},
numpages = {24}
}

@misc{planandsolve,
      title={Plan-and-Solve Prompting: Improving Zero-Shot Chain-of-Thought Reasoning by Large Language Models}, 
      author={Lei Wang and Wanyu Xu and Yihuai Lan and Zhiqiang Hu and Yunshi Lan and Roy Ka-Wei Lee and Ee-Peng Lim},
      year={2023},
      eprint={2305.04091},
      archivePrefix={arXiv},
      primaryClass={cs.CL},
      url={https://arxiv.org/abs/2305.04091}, 
}

@article{graphofthoughts,
   title={Graph of Thoughts: Solving Elaborate Problems with Large Language Models},
   volume={38},
   ISSN={2159-5399},
   url={http://dx.doi.org/10.1609/aaai.v38i16.29720},
   DOI={10.1609/aaai.v38i16.29720},
   number={16},
   journal={Proceedings of the AAAI Conference on Artificial Intelligence},
   publisher={Association for the Advancement of Artificial Intelligence (AAAI)},
   author={Besta, Maciej and Blach, Nils and Kubicek, Ales and Gerstenberger, Robert and Podstawski, Michal and Gianinazzi, Lukas and Gajda, Joanna and Lehmann, Tomasz and Niewiadomski, Hubert and Nyczyk, Piotr and Hoefler, Torsten},
   year={2024},
   month=mar, pages={17682–17690} }

@misc{treeofthoughts,
      title={Tree of Thoughts: Deliberate Problem Solving with Large Language Models}, 
      author={Shunyu Yao and Dian Yu and Jeffrey Zhao and Izhak Shafran and Thomas L. Griffiths and Yuan Cao and Karthik Narasimhan},
      year={2023},
      eprint={2305.10601},
      archivePrefix={arXiv},
      primaryClass={cs.CL},
      url={https://arxiv.org/abs/2305.10601}, 
}

@misc{rewoo,
      title={ReWOO: Decoupling Reasoning from Observations for Efficient Augmented Language Models}, 
      author={Binfeng Xu and Zhiyuan Peng and Bowen Lei and Subhabrata Mukherjee and Yuchen Liu and Dongkuan Xu},
      year={2023},
      eprint={2305.18323},
      archivePrefix={arXiv},
      primaryClass={cs.CL},
      url={https://arxiv.org/abs/2305.18323}, 
}

@misc{gorilla,
      title={Gorilla: Large Language Model Connected with Massive APIs}, 
      author={Shishir G. Patil and Tianjun Zhang and Xin Wang and Joseph E. Gonzalez},
      year={2023},
      eprint={2305.15334},
      archivePrefix={arXiv},
      primaryClass={cs.CL},
      url={https://arxiv.org/abs/2305.15334}, 
}

@misc{toolLLM,
      title={ToolLLM: Facilitating Large Language Models to Master 16000+ Real-world APIs}, 
      author={Yujia Qin and Shihao Liang and Yining Ye and Kunlun Zhu and Lan Yan and Yaxi Lu and Yankai Lin and Xin Cong and Xiangru Tang and Bill Qian and Sihan Zhao and Lauren Hong and Runchu Tian and Ruobing Xie and Jie Zhou and Mark Gerstein and Dahai Li and Zhiyuan Liu and Maosong Sun},
      year={2023},
      eprint={2307.16789},
      archivePrefix={arXiv},
      primaryClass={cs.AI},
      url={https://arxiv.org/abs/2307.16789}, 
}

@inproceedings{
toolformer,
title={Toolformer: Language Models Can Teach Themselves to Use Tools},
author={Timo Schick and Jane Dwivedi-Yu and Roberto Dessi and Roberta Raileanu and Maria Lomeli and Eric Hambro and Luke Zettlemoyer and Nicola Cancedda and Thomas Scialom},
booktitle={Thirty-seventh Conference on Neural Information Processing Systems},
year={2023},
url={https://openreview.net/forum?id=Yacmpz84TH}
}

@inproceedings{
metagpt,
title={Meta{GPT}: Meta Programming for A Multi-Agent Collaborative Framework},
author={Sirui Hong and Mingchen Zhuge and Jonathan Chen and Xiawu Zheng and Yuheng Cheng and Jinlin Wang and Ceyao Zhang and Zili Wang and Steven Ka Shing Yau and Zijuan Lin and Liyang Zhou and Chenyu Ran and Lingfeng Xiao and Chenglin Wu and J{\"u}rgen Schmidhuber},
booktitle={The Twelfth International Conference on Learning Representations},
year={2024},
url={https://openreview.net/forum?id=VtmBAGCN7o}
}

@misc{hugginggpt,
      title={HuggingGPT: Solving AI Tasks with ChatGPT and its Friends in Hugging Face}, 
      author={Yongliang Shen and Kaitao Song and Xu Tan and Dongsheng Li and Weiming Lu and Yueting Zhuang},
      year={2023},
      eprint={2303.17580},
      archivePrefix={arXiv},
      primaryClass={cs.CL},
      url={https://arxiv.org/abs/2303.17580}, 
}

@misc{agentxray,
      title={AgentXRay: White-Boxing Agentic Systems via Workflow Reconstruction}, 
      author={Ruijie Shi and Houbin Zhang and Yuecheng Han and Yuheng Wang and Jingru Fan and Runde Yang and Yufan Dang and Huatao Li and Dewen Liu and Yuan Cheng and Chen Qian},
      year={2026},
      eprint={2602.05353},
      archivePrefix={arXiv},
      primaryClass={cs.AI},
      url={https://arxiv.org/abs/2602.05353}, 
}

@inproceedings{
aflow,
title={{AF}low: Automating Agentic Workflow Generation},
author={Jiayi Zhang and Jinyu Xiang and Zhaoyang Yu and Fengwei Teng and Xiong-Hui Chen and Jiaqi Chen and Mingchen Zhuge and Xin Cheng and Sirui Hong and Jinlin Wang and Bingnan Zheng and Bang Liu and Yuyu Luo and Chenglin Wu},
booktitle={The Thirteenth International Conference on Learning Representations},
year={2025},
url={https://openreview.net/forum?id=z5uVAKwmjf}
}

@misc{irtrouter,
      title={IRT-Router: Effective and Interpretable Multi-LLM Routing via Item Response Theory}, 
      author={Wei Song and Zhenya Huang and Cheng Cheng and Weibo Gao and Bihan Xu and GuanHao Zhao and Fei Wang and Runze Wu},
      year={2025},
      eprint={2506.01048},
      archivePrefix={arXiv},
      primaryClass={cs.AI},
      url={https://arxiv.org/abs/2506.01048}, 
}

@misc{helium,
      title={Efficient LLM Serving for Agentic Workflows: A Data Systems Perspective}, 
      author={Noppanat Wadlom and Junyi Shen and Yao Lu},
      year={2026},
      eprint={2603.16104},
      archivePrefix={arXiv},
      primaryClass={cs.MA},
      url={https://arxiv.org/abs/2603.16104}, 
}

@misc{autellix,
      title={Autellix: An Efficient Serving Engine for LLM Agents as General Programs}, 
      author={Michael Luo and Xiaoxiang Shi and Colin Cai and Tianjun Zhang and Justin Wong and Yichuan Wang and Chi Wang and Yanping Huang and Zhifeng Chen and Joseph E. Gonzalez and Ion Stoica},
      year={2025},
      eprint={2502.13965},
      archivePrefix={arXiv},
      primaryClass={cs.LG},
      url={https://arxiv.org/abs/2502.13965}, 
}

@inproceedings{xgboost,
author = {Chen, Tianqi and Guestrin, Carlos},
title = {XGBoost: A Scalable Tree Boosting System},
year = {2016},
isbn = {9781450342322},
publisher = {Association for Computing Machinery},
address = {New York, NY, USA},
url = {https://doi.org/10.1145/2939672.2939785},
doi = {10.1145/2939672.2939785},
booktitle = {Proceedings of the 22nd ACM SIGKDD International Conference on Knowledge Discovery and Data Mining},
pages = {785–794},
numpages = {10},
location = {San Francisco, California, USA},
series = {KDD '16}
}

@misc{chaudhry2025murakkab,
      title={Murakkab: Resource-Efficient Agentic Workflow Orchestration in Cloud Platforms}, 
      author={Gohar Irfan Chaudhry and Esha Choukse and Haoran Qiu and Íñigo Goiri and Rodrigo Fonseca and Adam Belay and Ricardo Bianchini},
      year={2025},
      eprint={2508.18298},
      archivePrefix={arXiv},
      primaryClass={cs.MA},
      url={https://arxiv.org/abs/2508.18298}, 
}

@misc{madaan2023selfrefine,
    title={Self-Refine: Iterative Refinement with Self-Feedback}, 
    author={Aman Madaan and Niket Tandon and Prakhar Gupta and Skyler Hallinan and Luyu Gao and Sarah Wiegreffe and Uri Alon and Nouha Dziri and Shrimai Prabhumoye and Yiming Yang and Sean Welleck and Bodhisattwa Prasad Majumder and Shashank Gupta and Amir Yazdanbakhsh and Peter Clark},
    year={2023},
    eprint={2303.17651},
    archivePrefix={arXiv},
    primaryClass={cs.CL}
}

@misc{shinn2023reflexion,
      title={Reflexion: Language Agents with Verbal Reinforcement Learning}, 
      author={Noah Shinn and Federico Cassano and Edward Berman and Ashwin Gopinath and Karthik Narasimhan and Shunyu Yao},
      year={2023},
      eprint={2303.11366},
      archivePrefix={arXiv},
      primaryClass={cs.AI}
}

@article{
    chen2024frugalgpt,
    title={Frugal{GPT}: How to Use Large Language Models While Reducing Cost and Improving Performance},
    author={Lingjiao Chen and Matei Zaharia and James Zou},
    journal={Transactions on Machine Learning Research},
    issn={2835-8856},
    year={2024},
    url={https://openreview.net/forum?id=cSimKw5p6R},
    note={Featured Certification}
}

@inproceedings{
ong2025routellm,
title={Route{LLM}: Learning to Route {LLM}s from Preference Data},
author={Isaac Ong and Amjad Almahairi and Vincent Wu and Wei-Lin Chiang and Tianhao Wu and Joseph E. Gonzalez and M Waleed Kadous and Ion Stoica},
booktitle={The Thirteenth International Conference on Learning Representations},
year={2025},
url={https://openreview.net/forum?id=8sSqNntaMr}
}

@misc{ding2024hybridllm,
      title={Hybrid LLM: Cost-Efficient and Quality-Aware Query Routing}, 
      author={Dujian Ding and Ankur Mallick and Chi Wang and Robert Sim and Subhabrata Mukherjee and Victor Ruhle and Laks V. S. Lakshmanan and Ahmed Hassan Awadallah},
      year={2024},
      eprint={2404.14618},
      archivePrefix={arXiv},
      primaryClass={cs.LG},
      url={https://arxiv.org/abs/2404.14618}
}

@inproceedings{
tsiourvas2025causal,
title={Causal {LLM} Routing: End-to-End Regret Minimization from Observational Data},
author={Asterios Tsiourvas and Wei Sun and Georgia Perakis},
booktitle={The Thirty-ninth Annual Conference on Neural Information Processing Systems},
year={2025},
url={https://openreview.net/forum?id=iZC5xoQQkX}
}

@inproceedings{dekoninck2025unified,
  author={Jasper Dekoninck and Maximilian Baader and Martin T. Vechev},
  title={A Unified Approach to Routing and Cascading for LLMs},
  year={2025},
  cdate={1735689600000},
  url={https://proceedings.mlr.press/v267/dekoninck25a.html},
  booktitle={ICML},
}

@inproceedings {crankshaw2017clipper,
author = {Daniel Crankshaw and Xin Wang and Guilio Zhou and Michael J. Franklin and Joseph E. Gonzalez and Ion Stoica},
title = {Clipper: A {Low-Latency} Online Prediction Serving System},
booktitle = {14th USENIX Symposium on Networked Systems Design and Implementation (NSDI 17)},
year = {2017},
isbn = {978-1-931971-37-9},
address = {Boston, MA},
pages = {613--627},
url = {https://www.usenix.org/conference/nsdi17/technical-sessions/presentation/crankshaw},
publisher = {USENIX Association},
month = mar
}

@inproceedings {romero2021infaas,
author = {Francisco Romero and Qian Li and Neeraja J. Yadwadkar and Christos Kozyrakis},
title = {{INFaaS}: Automated Model-less Inference Serving},
booktitle = {2021 USENIX Annual Technical Conference (USENIX ATC 21)},
year = {2021},
isbn = {978-1-939133-23-6},
pages = {397--411},
url = {https://www.usenix.org/conference/atc21/presentation/romero},
publisher = {USENIX Association},
month = jul
}

@inproceedings{crankshaw2020inferline,
author = {Crankshaw, Daniel and Sela, Gur-Eyal and Mo, Xiangxi and Zumar, Corey and Stoica, Ion and Gonzalez, Joseph and Tumanov, Alexey},
title = {InferLine: latency-aware provisioning and scaling for prediction serving pipelines},
year = {2020},
isbn = {9781450381376},
publisher = {Association for Computing Machinery},
address = {New York, NY, USA},
url = {https://doi.org/10.1145/3419111.3421285},
doi = {10.1145/3419111.3421285},
booktitle = {Proceedings of the 11th ACM Symposium on Cloud Computing},
pages = {477–491},
numpages = {15},
location = {Virtual Event, USA},
series = {SoCC '20}
}

@inproceedings{romero2021llama,
author = {Romero, Francisco and Zhao, Mark and Yadwadkar, Neeraja J. and Kozyrakis, Christos},
title = {Llama: A Heterogeneous \& Serverless Framework for Auto-Tuning Video Analytics Pipelines},
year = {2021},
isbn = {9781450386388},
publisher = {Association for Computing Machinery},
address = {New York, NY, USA},
url = {https://doi.org/10.1145/3472883.3486972},
doi = {10.1145/3472883.3486972},
booktitle = {Proceedings of the ACM Symposium on Cloud Computing},
pages = {1–17},
numpages = {17},
location = {Seattle, WA, USA},
series = {SoCC '21}
}

@article{chung2025longcontext,
  title        = {Is Long Context All You Need? Leveraging {LLM}'s Extended Context for {NL2SQL}},
  author       = {Chung, Yeounoh and Kakkar, Gaurav T. and Gan, Yu and Milne, Brenton and Ozcan, Fatma},
  journal      = {arXiv preprint arXiv:2501.12372},
  year         = {2025},
  doi          = {10.48550/arXiv.2501.12372},
  eprint       = {2501.12372},
  archivePrefix= {arXiv},
  primaryClass = {cs.DB}
}

@article{renze2024selfreflection,
  title        = {Self-Reflection in {LLM} Agents: Effects on Problem-Solving Performance},
  author       = {Renze, Matthew and Guven, Erhan},
  journal      = {arXiv preprint arXiv:2405.06682},
  year         = {2024},
  doi          = {10.48550/arXiv.2405.06682},
  eprint       = {2405.06682},
  archivePrefix= {arXiv},
  primaryClass = {cs.CL}
}

@inproceedings{zheng2023sglang,
author = {Zheng, Lianmin and Yin, Liangsheng and Xie, Zhiqiang and Sun, Chuyue and Huang, Jeff and Yu, Cody Hao and Cao, Shiyi and Kozyrakis, Christos and Stoica, Ion and Gonzalez, Joseph E. and Barrett, Clark and Sheng, Ying},
title = {SGLang: efficient execution of structured language model programs},
year = {2024},
isbn = {9798331314385},
publisher = {Curran Associates Inc.},
address = {Red Hook, NY, USA},
booktitle = {Proceedings of the 38th International Conference on Neural Information Processing Systems},
articleno = {2000},
numpages = {27},
location = {Vancouver, BC, Canada},
series = {NIPS '24}
}

@article{pan2025kvflow,
  title = {KVFlow: Efficient Prefix Caching for Accelerating LLM-Based Multi-Agent Workflows},
  author = {Pan, Zaifeng and Patel, Ajjkumar and Hu, Zhengding and Shen, Yipeng and Guan, Yue and Li, Wan-Lu and Qin, Lianhui and Wang, Yida and Ding, Yufei},
  journal = {arXiv preprint arXiv:2507.07400},
  year = {2025}
}

@inproceedings{mallick2022matchmaker,
	author = {Mallick, Ankur and Hsieh, Kevin and Arzani, Behnaz and Joshi, Gauri},
	booktitle = {Proceedings of Machine Learning and Systems},
	editor = {D. Marculescu and Y. Chi and C. Wu},
	pages = {77--94},
	title = {Matchmaker: Data Drift Mitigation in Machine Learning for Large-Scale Systems},
	volume = {4},
	year = {2022}
}

@inproceedings{lewis2022augur,
author = {Lewis, Grace A. and Echeverr\'{\i}a, Sebasti\'{a}n and Pons, Lena and Chrabaszcz, Jeffrey},
title = {Augur: a step towards realistic drift detection in production ML systems},
year = {2023},
isbn = {9781450393195},
publisher = {Association for Computing Machinery},
address = {New York, NY, USA},
url = {https://doi.org/10.1145/3526073.3527590},
doi = {10.1145/3526073.3527590},
booktitle = {Proceedings of the 1st Workshop on Software Engineering for Responsible AI},
pages = {37–44},
numpages = {8},
location = {Pittsburgh, Pennsylvania},
series = {SE4RAI '22}
}

@misc{langgraph,
    title = {LangGraph},
    author = {{LangChain, Inc.}},
    year = {2024},
    url = {https://github.com/langchain-ai/langgraph}
}

@misc{llamaindex,
author = {Liu, Jerry},
doi = {10.5281/zenodo.1234},
month = {11},
title = {{LlamaIndex}},
url = {https://github.com/run-llama/llama_index},
year = {2022}
}

@misc{ag2,
    title = {AutoGen: Enabling Next-Gen {LLM} Applications via Multi-Agent Conversation Framework},
    author = {Qingyun Wu and Gagan Bansal and Jieyu Zhang and Yiran Wu and Beibin Li and Eric Zhu and Li Jiang and Shaokun Zhang and Xiaoyun Zhang and Jiale Liu and Ahmed Hassan Awadallah and Ryen W. White and Doug Burger and Chi Wang},
    year = {2023},
    eprint = {2308.08155},
    archivePrefix = {arXiv},
    primaryClass = {cs.AI}
}

@article{candes2009matrixcompletion,
  author    = {Cand{\`e}s, Emmanuel J. and Recht, Benjamin},
  title     = {Exact Matrix Completion via Convex Optimization},
  journal   = {Foundations of Computational Mathematics},
  volume    = {9},
  number    = {6},
  pages     = {717--772},
  year      = {2009},
}

@article{koren2009matrixfactorization,
  author    = {Koren, Yehuda and Bell, Robert and Volinsky, Chris},
  title     = {Matrix Factorization Techniques for Recommender Systems},
  journal   = {Computer},
  volume    = {42},
  number    = {8},
  pages     = {30--37},
  year      = {2009},
  doi       = {10.1109/MC.2009.263},
}

@article{Dspy,
  title={Dspy: Compiling declarative language model calls into self-improving pipelines},
  author={Khattab, Omar and Singhvi, Arnav and Maheshwari, Paridhi and Zhang, Zhiyuan and Santhanam, Keshav and Vardhamanan, Sri and Haq, Saiful and Sharma, Ashutosh and Joshi, Thomas T and Moazam, Hanna and others},
  journal={arXiv preprint arXiv:2310.03714},
  year={2023}
}

\newpage
\appendix

\section{Estimating Accuracy for Workflow Instances under MNAR Sampling}
\label{sec:appendix:matrix}

This appendix describes \sys's approach to estimating the expected accuracy of
every workflow instance from sparse offline profiling.  We show that the cascade
structure of the workflow tree, which at first appears to make the missing-data
problem harder, in fact provides an exact bias correction that standard matrix
completion cannot exploit.

\subsection{Problem Statement}
\label{sec:appendix:matrix:setup}

Let $\QuerySet = \{q_1, \ldots, q_m\}$ be the set of profiling queries and let
$\mathcal{P}$ denote the set of all workflow instances (\ie root-to-leaf paths
in the workflow tree).  The accuracy outcomes form a request--path table
$A \in \{0,1\}^{|\QuerySet| \times |\mathcal{P}|}$, where $A(q, p) = 1$ if
workflow instance $p$ succeeds on query $q$.  The goal of offline profiling
is to estimate the \emph{expected accuracy} of each workflow instance:
\begin{equation}
  \mu(p) = \frac{1}{|\QuerySet|}\sum_{q \in \QuerySet} A(q,p)
         = \mathbb{E}_{q}[A(q,p)], \quad p \in \mathcal{P}.
  \label{eq:colmean}
\end{equation}
Given a cost or latency budget that limits the number of profiling runs, only a
small fraction of entries in $A$ can be observed directly.  Estimating $\mu$
from this partial view is the core matrix-completion problem.

\subsection{The MNAR Challenge}
\label{sec:appendix:matrix:mnar}

Cascade sampling is the natural way to profile the workflow tree: pick a query
$q$ uniformly at random, invoke a randomly sampled depth-1 model $L_i$, and
continue to deeper stages only when the current execution fails.  Under a
profiling budget of $N$ runs, this produces $N$ observed entries distributed
across $A$, but the observation pattern is
\textbf{Missing Not At Random (MNAR)}: a depth-2 workflow instance
$(L_i, L_j)$ is only observed on queries for which $L_i$ already failed.
Formally, for any query $q$,
\begin{equation}
\begin{split}
  \mathbb{E}\bigl[A(q,\,(L_i, L_j)) \mid (q,\,(L_i,L_j)) \in \Omega\bigr]
  &= q(L_j \mid L_i\text{ fails}) \\
  &\neq \mu(L_i, L_j),
\end{split}
\label{eq:mnar}
\end{equation}
where $\Omega$ is the set of observed entries and
$q(L_j \mid L_i\text{ fails})$ is the conditional accuracy of $L_j$ given that
$L_i$ already failed.  The two quantities differ because $\mu(L_i, L_j)$
averages over \emph{all} queries, while the observed entries come from the
harder subpopulation on which $L_i$ failed.  Depth-3 instances are even more
severely affected: they are observed only on the subset of queries where both
depth-1 and depth-2 stages fail, \ie the hardest queries.

Standard matrix completion applies low-rank factorization to the observed
entries, fitting the latent model to this biased subpopulation. This yields large errors for
depth-2 and depth-3 instances even after subtree fill-in.

\subsection{Unbiased Estimation via Cascade Decomposition}
\label{sec:appendix:matrix:decomp}

The key insight is that the cascade workflow structure, despite inducing MNAR,
also reveals its own bias correction.  Because $A$ is prefix-closed --- a
workflow instance succeeds whenever any of its prefixes succeed --- the
expected accuracy satisfies the following recursive decomposition:
\begin{align}
  \mu(L_i) &= \mathbb{E}_{q}[A(q, L_i)], \label{eq:d1}\\
  \mu(L_i, L_j) &= \mu(L_i)
    + \bigl(1 - \mu(L_i)\bigr)\cdot q(L_j \mid L_i\text{ fails}), \label{eq:d2}\\
  \mu(L_i,L_j,L_k) &= \mu(L_i,L_j)
    + \bigl(1-\mu(L_i,L_j)\bigr)\cdot q(L_k \mid (L_i,L_j)\text{ fails}). \label{eq:d3}
\end{align}
The conditional accuracy $q(L_j \mid L_i\text{ fails})$ in \eqref{eq:d2} is
precisely the quantity that \eqref{eq:mnar} shows to be \emph{unbiasedly}
estimated by the sample mean of observed entries in column $(L_i, L_j)$.
Substituting sample means into \eqref{eq:d1}--\eqref{eq:d3} therefore yields
a consistent estimator:
\begin{align}
  \hat{\mu}(L_i) &= \bar{A}_{L_i},
    \label{eq:est-d1}\\
  \hat{\mu}(L_i,L_j) &= \hat{\mu}(L_i)
    + \bigl(1-\hat{\mu}(L_i)\bigr)\cdot \bar{A}_{(L_i,L_j)},
    \label{eq:est-d2}\\
  \hat{\mu}(L_i,L_j,L_k) &= \hat{\mu}(L_i,L_j)
    + \bigl(1-\hat{\mu}(L_i,L_j)\bigr)\cdot \bar{A}_{(L_i,L_j,L_k)},
    \label{eq:est-d3}
\end{align}
where $\bar{A}_p$ denotes the sample mean of observed entries in column $p$.
As $N \to \infty$, $\hat{\mu}(p) \to \mu(p)$ for every workflow instance $p$.

\heading{Intuition}
The MNAR observation model is harmful for estimating $\mu(p)$ directly, but
it is exactly the right conditioning for estimating the incremental conditional
accuracy $q$.  The cascade decomposition turns this into an advantage: rather
than estimating $\mu$ from biased column means, we estimate $q$ from unbiased
conditional means and reconstruct $\mu$ via the recursive formula.

\subsection{Low-Rank Regularisation for Sparse Depth-3 Estimates}
\label{sec:appendix:matrix:svd}

The estimator in \eqref{eq:est-d1}--\eqref{eq:est-d3} is unbiased but
high-variance for depth-3 instances: with a 5\% profiling budget, depth-3
columns receive only ${\approx}20$--$80$ observations each.

To reduce this variance, we exploit the low-rank structure of the
\emph{conditional accuracy matrix} $Q$.  Define the $72 \times |\LLMSet|$
matrix $Q$ whose rows correspond to every workflow instance prefix
(8 depth-1 prefixes and 64 depth-2 prefixes) and whose columns correspond to
candidate last-stage models, with $Q[p, L_k] = q(L_k \mid p\text{ fails})$.
The depth-3 block (rows corresponding to depth-2 prefixes) is approximately
rank-1 in practice, reflecting a single latent query-difficulty axis that
drives conditional accuracy across all prefixes and models.

We apply a rank-1 SVD projection to the $64 \times |\LLMSet|$ depth-3 block:
\begin{equation}
  \hat{Q}_{\text{depth-3}} = \Pi_1\!\left(Q_{\text{depth-3}}\right),
  \qquad
  \hat{Q}_{\text{depth-2}} = Q_{\text{depth-2}},
  \label{eq:svd}
\end{equation}
where $\Pi_1$ denotes projection onto the rank-1 manifold (initialized with
column means for unobserved entries).  SVD is applied only to the sparse
depth-3 block; the depth-2 block is well-observed (all 64 entries, 150--500
samples each) and uses raw conditional means directly to avoid introducing
bias.  The smoothed rates $\hat{Q}$ are then substituted into
\eqref{eq:est-d2}--\eqref{eq:est-d3} in place of $\bar{A}_p$.

\end{document}